\documentclass[preprint, 12pt,review,authoryear]{elsarticle}
\usepackage[margin = 3cm]{geometry}
\usepackage{times}
\usepackage{multirow}
\usepackage{graphicx}
\usepackage{lscape}
\usepackage{caption}
\usepackage{subcaption}
\usepackage{amssymb}
\usepackage{amsmath}
\usepackage{multirow}
\usepackage{lineno}
\usepackage{booktabs}
\usepackage{footnote}
\usepackage{setspace}
\usepackage{natbib}
\usepackage{color}
\usepackage{wasysym}
\usepackage[table,xcdraw]{xcolor}
\usepackage{threeparttable}
\usepackage{textcomp}
\usepackage{tfrupee}
\usepackage[toc,page]{appendix}
\usepackage[utf8]{inputenc}
\usepackage{newunicodechar}
\newunicodechar{²}{\ensuremath{{}^2}}
\usepackage{csquotes}

\makeatletter
\renewcommand{\fnum@figure}{Fig. \thefigure}
\makeatother

\begin{document}

\begin{frontmatter}


\title{Impact of consumer preferences on decarbonization of transport sector in India}



\author{Nandita Saraf}
\author{Yogendra Shastri}

\address{Department of Chemical engineering, Indian Institute of Technology, Bombay, Mumbai, India}

\begin{abstract}
 
Decarbonization of transport sector through adoption of cleaner vehicle options will depend on the environmental awareness of consumers and their priorities. This work develops and uses a system dynamics approach to understand possible adoption pathways of novel vehicle options, i.e., ethanol-blended fuel (E85), electric, and compressed natural gas (CNG) vehicles in India. A system dynamics model using a multi-nominal logit model to capture consumer choices has been previously developed for the private road transport sector of India. The model has been modified to also include annual vehicular emissions as a decision-making attribute in addition to the annual cost. Four different classes of consumers with different priorities to cost and annual emissions are modeled. The model coefficients are identified using historical data. Model simulations over a period of 30 years till 2050 are performed to determine possible vehicle adoption trends. Different scenarios of changing environmental awareness, policy interventions, and technology development were analyzed to achieve targets such as COP26 greenhouse gas emission reduction, ethanol blending, and EV adoption. Simulation results showed that an increase in environmental awareness resulted in the adoption of novel vehicle options by 67.42\% and 22.3\% of the car and two-wheeler stocks, respectively. However, rising environmental awareness was not enough to meet the target values. Scenario analysis showed that a greater share of renewables in the electricity grid, carbon tax on transport fuel, and reduced vehicle driving together can possibly achieve the GHG emissions target in 2030. Battery-related GHG emissions were shown to be very important and led to counter-intuitive trends in EV adoption. 30\% EV sales by 2030 could be achieved with a greener electricity grid and carbon tax. The model can be used as decision support to test various policy alternatives and provide quantitative estimates.

\end{abstract}


\end{frontmatter}

%%\begin{frontmatter}



\section{Introduction}
\label{S:1}

At the COP26 summit, India pledged to reduce the emission intensity by 45\% over the 2005 level by 2030 and become a net zero carbon emitter by 2070. Strategies to achieve these targets will require focusing on India's emission-intensive sectors. The transport sector contributes to about 10\% of India's total greenhouse gas(GHG) emission, making it the second most carbon-emitting sector in the country \citep{terii}. Among various modes of transport, road transport has seen the highest growth in travel demand by 11\% between 2000 and 2012, due to increased private vehicle ownership, i.e., cars and two-wheelers \citep{singh2019greenhouse}. Consequently, the road transport sector emits about 90-92\% of the total transport sector's GHG emissions \citep{ceeww}. With continued population and per capita income growth, the private transport sector is expected to witness an exponential growth in vehicle and travel demand, causing increased energy consumption and emissions \citep{dhar2018transformation}. To reduce the consumption of carbon-intensive fossil fuels and therefore GHG emissions, a number of novel private vehicle options expected to be more sustainable, are being promoted by the Indian government through policies and incentives. Adoption of E85 vehicles (driven on the blend of 85\% ethanol in 15\% of petrol) is being encouraged to achieve 20\% ethanol blending in petrol by 2025 as per National Biofuels Policy \citep{niti}. Central and state-level subsidies and incentives are being provided for the purchase of electric vehicles (EVs) to reduce their ownership cost and achieve 30\% EV sales in the new sales by 2030 as per EV30@30 policy. An initiative \enquote{Sustainable Alternative Towards Affordable Transportation} (SATAT) has been undertaken by the government to extract economic value from biomass waste by converting it to compressed biogas (CBG) and bio-manure. CBG will replace the fossil compressed natural gas (CNG) used in the transport sector. SATAT scheme has targeted to set up 5000 compressed biogas (CBG) plants by 2023 to produce 15MMT of CBG that could be used to fuel CNG vehicles. Additionally, the government is encouraging the adoption of low-emission options by raising awareness about climate change and its adverse effects. 

Despite these initiatives, the adoption rate of these new vehicle options is low. One of the reasons is the higher purchase or operating cost of these vehicles, as well as the lack of related infrastructure such as EV charging stations. \cite{saraf2021development} developed a system dynamics model to understand the adoption of novel transport options as a function of the annual cost of various vehicle options. The multi nominal logit model was used to formulate the purchase decision, which was solely based on the ownership cost of the vehicle. Dynamics and feedback effects associated with fuel prices (petrol, diesel, and ethanol) as well as infrastructure development were incorporated. Their results showed that the targets set by the Government of India were too ambitious.

One of the limitations of their work was the consideration of only annual cost as a factor in consumer choice for selecting a vehicle option. In practice, consumer awareness about the environmental impact of the vehicles and the consumer’s prioritization of this impact relative to cost will also play a role in decision-making. Consumers are likely to become more aware of greener transport options, thereby pushing the adoption of novel options. According to the finding of Deloitte's 2020 Global Automotive Consumers Study, about 25\% of Indian consumers give the highest preference to hybrid EV, followed by 15\% to battery EV and 9\% to ethanol and CNG options \citep{survey2}. About 53\% of Indian consumers are also willing to pay more than one lakh to buy an EV. Another survey revealed that about 90\% of Indian consumers are willing to pay a premium to buy an EV \citep{survey1}. These findings indicate that environmental awareness is becoming important in the purchase decisions of Indian consumers. Therefore, including these factors in decision-making models is important. The insights obtained through such an assessment can be used by the government to plan targeted awareness campaigns. The broad objective of this work is to incorporate these factors in the system dynamics model previously developed, and provide more realistic insights.

Agent-based modeling (ABM) approach has extensive use in the study of multi-level interaction of individual behavior and social dynamics. A particular application of ABM involves innovation diffusion, specifically in the area of environmental innovation like the adoption of alternative vehicle options in the transport sector \citep{shafiei2015simulation,noori2016development}. Consumers are the prime agents in the adoption of new vehicle technologies. Often, the purchase decisions made by consumers are influenced by multiple attributes of the vehicles. These attributes can be classified as financial, technical, infrastructural, and policy attributes \citep{liao2017consumer}. Some of the vehicle attributes under these categories are purchase price, operating cost, driving range, charging time, charging station availability, and government incentives\citep{hackbarth2013consumer,yan2019research,bansal2021willingness,ghasri2019perception,sierzchula2014influence}. Several studies in this field employ discrete choice models such as standard logistic regression \citep{bjerkan2016incentives,mersky2016effectiveness,jansson2017adoption} or multi nominal models \citep{higgins2017size,krause2016assessing} to model the purchase decisions. \cite{noori2016development} coupled the ABM with multi nominal logit model (MNLM) to identify the market share of EVs in the US. The authors found that social acceptability or word-of-mouth will have a significant effect on EV market share. Social acceptability combined with government subsidies may achieve a 30\% EV share in new sales by 2030. A similar study was undertaken by \cite{hackbarth2013consumer} in Germany. Major findings revealed that German car buyers are very reluctant to buy battery and fuel cell-driven EVs. However, the plug-in hybrid EV has a higher chance of getting selected, especially among younger, highly educated, and environmentally conscious consumers. On the other hand, Canadian consumers are unwilling to pay the premium on EVs without EV experience  \citep{larson2014consumer}. A review of China's Electric Vehicle Subsidy Scheme revealed that financial incentives, as well as technology improvement associated with battery cost reduction, can make EV cost competitive \citep{hao2014china}. 

The literature on the impact of consumer preferences on vehicle selection is very limited in the Indian context. The first study in this regard was done by \cite{bansal2021willingness}. The study focused on understanding the effect of environmental friendliness and social norms on the likelihood of Indian car buyers adopting electric vehicles. Results showed that Indian consumers are willing to pay more to buy EVs with improved attributes i.e., more driving range, reduced charging time, and operating cost. Such studies contribute greatly to identifying the key barriers to the adoption of new vehicle options, devising informed market strategies based on consumer behavior, and optimizing vehicle design and cost to make vehicles affordable. Another study was conducted across India on consumer preference for electric two-wheeler over petrol two-wheeler, using a mixed logit model \citep{chakraborty2021acceptance}. Authors found that more environmentally concerned people and those who have greater knowledge of government policies and incentives on EVs have shown greater acceptance of electric two-wheelers. A system dynamics modelling approach was used by \cite{kushwah2021electric} to develop a causal loop diagram representing the underlying causal relations among various variables of the system which affect EV adoption in India. All these studies have provided useful information about factors that will dominate the decision-making process with regard to EV purchases. 

However, considering only electric vehicle adoption provides limited insight since other vehicle options such as, i.e., E85 and CNG are also commercially available in the market. Competition among these options will also affect consumer choice. Additionally, these studies do not give any quantitative information on EV adoption, i.e., the fraction of EV sales in the total sales in a particular time frame. This limits the ability to use results from these studies in an effective manner.

Therefore, extending the broad objective of this work stated previously, this work specifically aims to extend the SD framework developed by Saraf and Shastri (2022) so as to model varied consumer preferences for the selection of vehicle options including E85, EV, and CNG. A novel methodology has been developed to acknowledge different consumers and capture the diversity in consumer choices. Additionally, the impact of changing awareness with time is included. The research evaluates the vehicle adoption trends in the private road transport sector of India spanning a period of 30 years from 2020 to 2050. The research also evaluates different awareness scenarios with policy alternatives and technology development to achieve targets proposed by the government regarding ethanol blending and EV adoption. The work also quantifies the impact of adoption on GHG emissions and reports the likelihood of meeting the COP26 emission target for the private road transport sector by 2030. The impact of different feedback loops on the adoption trend and proposed targets are reported to emphasize the importance of considering the feedback effects of the system. Important contributions and novelty of this work are summarised here:
\begin{itemize}
        \item The model formulates the purchase decision of consumers based on ownership cost and annual emissions of the vehicle using the multi nominal logit model.
        \item The life cycle emission (LCE) of fuels, electricity grid, and manufacturing, recycling, and discarding of EV batteries have been used to estimate the annual emission of the vehicle.
        \item The dynamic features of the system affecting the ownership cost and annual emissions of the vehicle, i.e., the technological and infrastructural developments, supply-demand dynamics of fuels, and changing electricity grid have been considered in the model. These factors affect the adoption driven by ownership costs and annual emissions.  
        \item Differences in consumer preferences for ownership cost and annual emission of the vehicle have been captured in the model. Change in adoption trends as environmental awareness increases is also evaluated through scenarios.
        
\end{itemize}

The article is organised as follows. Section \ref{overview} provides the overview of the system dynamics model developed previously. Section \ref{modelformulation} explains the additions to the model to capture consumer choices and diversity. Parameterization of the model and description of various scenarios are discussed in section \ref{parameterization}. Results of these scenarios as well as implications of various feedback loops in the model are presented in section \ref{results}. The article concludes in section \ref{conclusion} followed by an outline of the future scope of this work.

\section{Overview of previous work}
\label{overview}

The transport sector is a complex dynamic system, consisting of a large number of variables which are constantly interacting with each other. It involves multiple stakeholders such as government, policymakers, consumers, oil marketing companies etc. The system dynamics (SD) model accepts complexity, non-linearity, feedback loop structures, and time delays that affect the system’s behaviour over time. Thus, it uses quantitative means to investigate the dynamic behaviour of the system. Therefore, the system dynamics methodology is best suited for understanding the behaviour of the transport sector.

A system dynamics model was developed by \cite{saraf2021development} to understand the adoption trends of various vehicle options, i.e., E85, EV, and CNG in India for a period of 30 years, between 2020 to 2050. The model only considers only private vehicle ownership, i.e., cars and two-wheelers. The model inputs are the annual demand for cars and two-wheelers, their respective travel demands, vehicle performance and purchase prices, life cycle emission, and the infrastructure development rates.

The model is divided into five sections, corresponding to each vehicle option, i.e., petrol, diesel, E85, electric, and CNG. Demand for cars is divided among five options, i.e., petrol, diesel, E85, electric, and CNG. Demand for two-wheelers is divided among three options, i.e., petrol, E85, and electric. The model is formulated on the assumption that the annual demand for cars and two-wheelers is divided among the available options based on the ownership cost of the vehicle. Higher ownership cost of a vehicle will reduce its annual demand and vice versa. The ownership cost of the vehicle is the sum of the annual vehicle purchase price, annual fuel expense, and annual maintenance cost. The fuel expense of the vehicle is affected by fuel price, which in turn is influenced by its demand and supply. Higher fuel demand will increase the fuel price whereas, a higher fuel supply will reduce the fuel price. Such dynamics have been captured in the model for petrol, diesel, and ethanol to determine their respective prices. Based on the respective fuel prices, the fuel expense, ownership cost, and annual demand of the vehicle for the next time step are estimated, thus forming negative feedback loops for petrol, diesel, and E85 vehicles. The new demand for each option alters the respective vehicle stock, which gives the measure of fuel demand. Molasses and lignocellulosic biomass have been considered feedstocks for ethanol production. Experience in process technology grows as production capacity expands, therefore production capacities of ethanol for both feedstocks positively affect the production cost. As production cost reduces, it increases the profit earned, which in turn will increase the production capacity, thus forming positive feedback loops. Since EV and CNG vehicles are evolving technologies, the lack of sufficient charging and gas-refilling stations poses a barrier to the adoption of these vehicle options. The lack of sufficient stations results in inconvenience which is measured by comparing the ideal and actual/existing number of stations. The ideal number of stations is the number required to refuel/charge the vehicle in a reasonable time. This inconvenience is converted to monetary terms and added to the ownership cost of the respective vehicles. Negative feedback loops have been identified connecting variables for respective vehicle stock, the ideal number of refueling stations, inconvenience cost, ownership cost of the vehicle, and demand for the vehicle. For further details on model formulations, the readers can refer to \cite{saraf2021development}.

The major limitations in the model are listed as follows: 

\begin{itemize}
    \item The model is formulated with the assumption that only the ownership cost of the vehicle affects its purchase decision.
    \item Although the inconvenience due to lack of sufficient refueling infrastructure and long refueling time is captured in the model in the form of inconvenience costs, the environmental performance of the vehicle is not considered.
    \item The model assumes a generic consumer behavior towards different vehicle options. It considers that all consumers are the same in terms of decision-making.
\end{itemize}

These limitations rule out the possibility that purchase decisions are also affected by factors like environmental awareness, reliability of the technology, brand loyalty, and societal status. In addition to that, the consumers in reality are independent agents with individual sets of rules for decision-making. Different combinations of factors affect their purchase decisions to different extents. Thus, to capture realistic consumer preferences for various options, i.e., E85, EV, and CNG vehicles and to address the above-mentioned limitations, this work has tried to model the purchase decision of consumers as a function of ownership cost and environmental awareness. Different kinds of consumers are expected to exhibit differences in vehicle selection criteria. For instance, a group of consumers may give more importance to economic factors than the environment, while the other group may give equal or no importance to economic and environmental factors, etc. Thus, a methodology has been developed to incorporate such differences in consumer preferences. This methodology also captures the change in consumer preferences as environmentally aware consumer increases in the future. The following section explains these modifications in detail.

\section{Model formulation}
\label{modelformulation}

This section first explains the inclusion of the environmental impact of a vehicle as an additional decision-making criterion in the multi-nominal logit model, followed by the description of the methodology to incorporate different consumer preferences in the model. 

\subsection{Modified multi-nominal logit model}

\begin{figure}
    \centering
    \includegraphics[width=17cm,height=17cm, keepaspectratio]{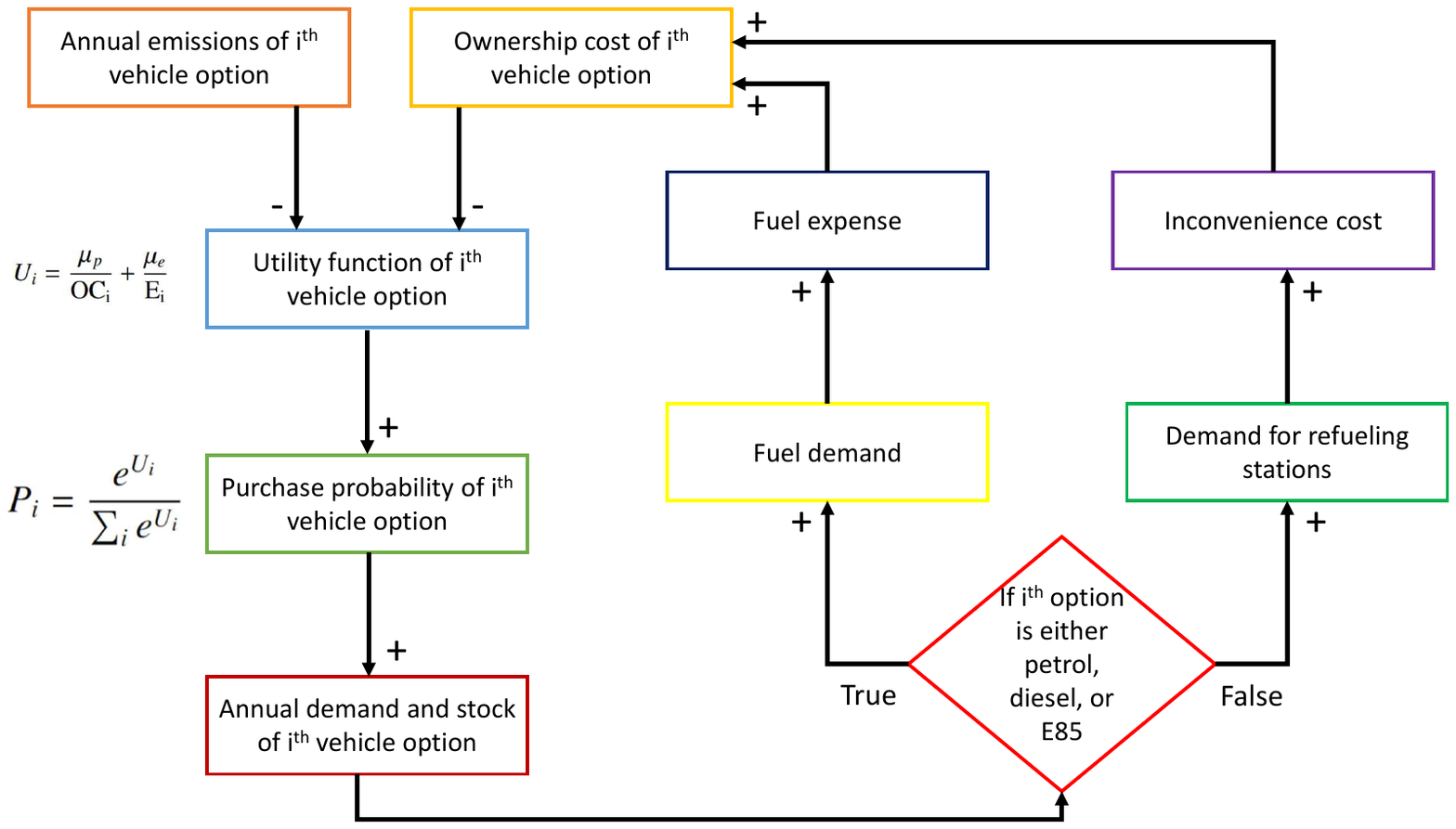}
    \caption{Schematic diagram to explain the multi-nominal logit model}
    \label{purchase}
\end{figure}

The general form of the multi-nominal logit model is given by the following Eq. \ref{generalprobability}.

\begin{equation}
    P_i = \frac{e^{U_i}}{\sum_{i}e^{U_i}}
    \label{generalprobability}
\end{equation}

\noindent where, $P_i$ is the purchase probability of $i^{th}$ vehicle option, $U_i$ is the utility function of the $i^{th}$ vehicle option. \cite{saraf2021development} considered annual cost adjusted for the inconvenience as the only factor in the calculation of the utility function. However, as mentioned previously, environmentally conscious consumers will also consider the emissions of the vehicle as an additional factor in their purchase decisions. The higher the environmental awareness among consumers, the more the low-emission vehicle options will be preferred. Therefore, the multi-nominal logit model has been modified in this work by considering the annual greenhouse gas emissions from the vehicle in the utility function. In the modified multi-nominal logit model, the utility function for each vehicle option is calculated as follows: 

\begin{equation}
    U_i =  \frac{\mu_p}{\mathrm{OC_i}} + \frac{\mu_e}{\mathrm{E_i}}
    \label{utility}
\end{equation}

\noindent where, $OC_i$ and $E_i$ are the ownership cost and annual emissions of $i^{th}$ vehicle option, $\mu_p$ and $\mu_e$ are the model parameters. The sensitivity of purchase probability to ownership cost is represented by $\mu_p$ while the sensitivity to annual emissions is represented by $\mu_e$. The rationale behind the formulation of Eq. \ref{generalprobability} and \ref{utility} is shown in Fig. \ref{purchase}. The ownership cost and annual emissions of the vehicle negatively impact the purchase probability (see Fig. \ref{purchase}), indicating an inverse relationship. Therefore, they are placed in the denominator of their respective terms in the utility function. Higher ownership cost and annual emissions will lower the value for $U_i$ in Eq. \ref{utility}, ultimately reducing the P, the purchase probability of the vehicle. The sensitivities of purchase probability with respect to the ownership cost and annual emissions can be varied by varying the $\mu_p$ and $\mu_e$ values. A High value of $\mu_p$ and a low value of $\mu_e$ make the first term in the utility function larger than the second term. Thus, the $U_i$ becomes more sensitive to the ownership cost of the vehicle. In contrast, a high value of $\mu_e$ and a low value of $\mu_p$ make the $U_i$ more sensitive to vehicle emissions. This feature can be used to capture the diversity of decision-makers in vehicle owners. 

Once the purchase probability is known from the value of $U_i$, the remaining causal relationships are the same as those explained by \cite{saraf2021development}. This includes negative feedback look on the fuel price and ownership cost of petrol, diesel, and E85 vehicles. EV or CNG/CBG vehicles, increased vehicle stock will increase the demand for refueling stations. This will increase the inconvenience cost and thereby the ownership cost of the vehicle. Similar to petrol, diesel, and E85, negative feedback loops are also identified for EV and CNG vehicle options. Here, the annual emissions are limited to greenhouse gases (GHGs) and are calculated from a life cycle perspective. Therefore, they include not only the emissions during the use phase of the fuel but also during the production phase. Annual emission of an electric vehicle includes GHG emission from the electricity grid and GHG emission during battery manufacturing, recycling, and discard processes. The next section describes how different types of consumers are modeled within the framework. 

\subsection{Modeling of consumer types}
\label{inclusion}

\begin{figure}[!ht]
    \centering
    \includegraphics[width=17cm,height=17cm, keepaspectratio]{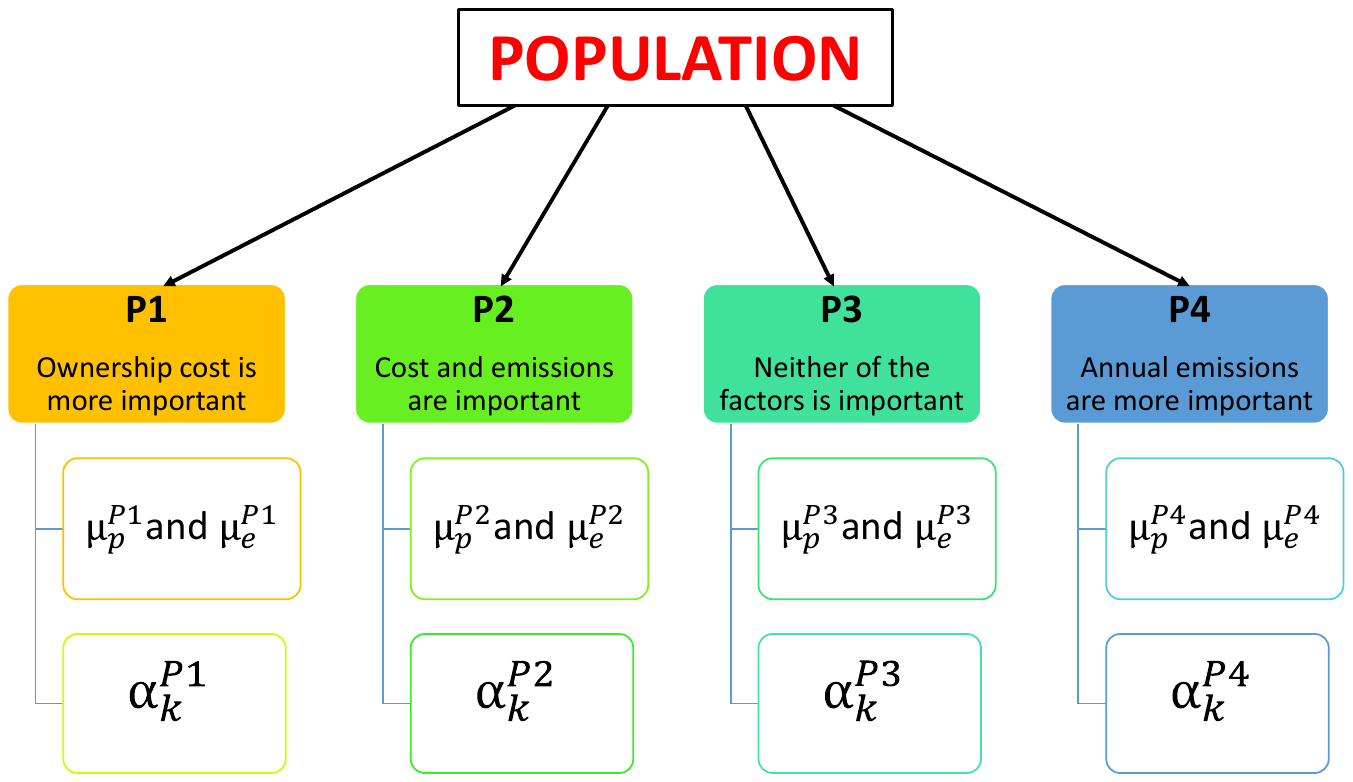}
    \caption{The schematic diagram demonstrates a methodology to incorporate the variations in preferences for different consumers}
    \label{population}
\end{figure}

In order to model diversity in consumer choices while purchasing new vehicles, two factors are important. First, different categories of consumers with differing priorities for annual cost and emissions need to be captured. Further, the distribution of the consumers among these different categories needs to be fixed. Fig. \ref{population} shows a schematic representation of the methodology adopted in this work to capture both these factors. 

It is possible to identify a large number of categories for consumers with varying priorities while making vehicle purchase decisions. It is difficult to consider all the kinds of consumers in the model. Therefore, to capture the variation in purchase decisions and also keep the model relatively simple, the consumers are broadly divided into four categories as follows: 

\begin{itemize}
    \item P1 is the group of consumers that considers the ownership cost as the most important factor in its purchase decision. These consumers prefer the low-cost vehicle option irrespective of its annual emissions.
    \item P2 is the group of consumers that gives equal importance to the ownership cost and annual emissions in its purchase decisions.
    \item Consumers who neither consider ownership cost nor the emissions in their purchase decision fall in the P3 category. They make decisions based on other factors such as brand loyalty, social status, or specific performance requirement.
    \item Environmentally aware consumers who give more importance to the annual emissions of the vehicle as compared to the annual cost in the purchase decision are categorized as P4. The ownership cost of the vehicle doesn't affect their purchase decisions.
\end{itemize}

For each of these four categories, the values of $\mu_p$ and $\mu_e$ differ, thereby capturing the different purchase priorities. For the P1 category, a high value for $\mu_p$ and a low value for $\mu_e$ are considered. For the P2 category, both $\mu_p$ and $\mu_e$ values are such that both governing factors will impact the purchase decision almost equally. When both $\mu_p$ and $\mu_e$ values are very low, neither of the factors affects the purchase decision, which is the case for the P3 category. For environmentally aware consumers in the P4 category, $\mu_e$ value is high while $\mu_p$ is low. The $\mu_p$ and $\mu_e$ values for each consumer category is designated by the symbols $\mu_p^{Pj}$ and $\mu_e^{Pj}$, where $Pj$ represents the consumer category and $j= {1,2,3,4}$ as shown in Fig. \ref{population}. For each category, the purchase probabilities for $i^{th}$ vehicle option for time step $k$ are termed as $P_{i_{k}}^{Pj}$. 

Once the categories are finalized, the consumers need to be divided among each of these categories. This is represented by $\alpha^{Pj}_k$, the fraction of consumer at time step falling in category $Pj$. This factor has been made a function of time step $k$ because the consumer's priorities may shift with time. This may be due to the following reasons or actions:

\begin{itemize}
    \item The global push on the government to meet emission reduction targets may lead to the prioritization of low-emission technologies.
    \item Per capita income may increase in the future, which may reduce the sensitivity of ownership cost in the purchase decision.
    \item Government's efforts to spread awareness about climate change and its adverse effects among the public through organizing campaigns, protests or rallies, spreading news through social media, etc. 
\end{itemize}

The proposed modeling approach allows capturing this shift in consumer preference or priorities with time. The weighted average purchase probability of $i^{th}$ vehicle option for time step $k$ is calculated using Eq. \ref{weightedavg}. 

\begin{equation}
    P_{i_{k}} = \sum_{j=1}^{4}\alpha _{k}^{Pj}\times P_{i_{k}}^{Pj}
    \label{weightedavg}
\end{equation}

Estimation of model parameters $\mu_p^{Pj}$, $\mu_e^{Pj}$, and $\alpha^{Pj}_k$ for the model is explained later as part of the model parameterization section.   

\subsection{Modified causal loop diagram}
\label{cl}

\begin{figure}
    \centering
    \includegraphics[width=19cm,height=19cm,angle=90, keepaspectratio]{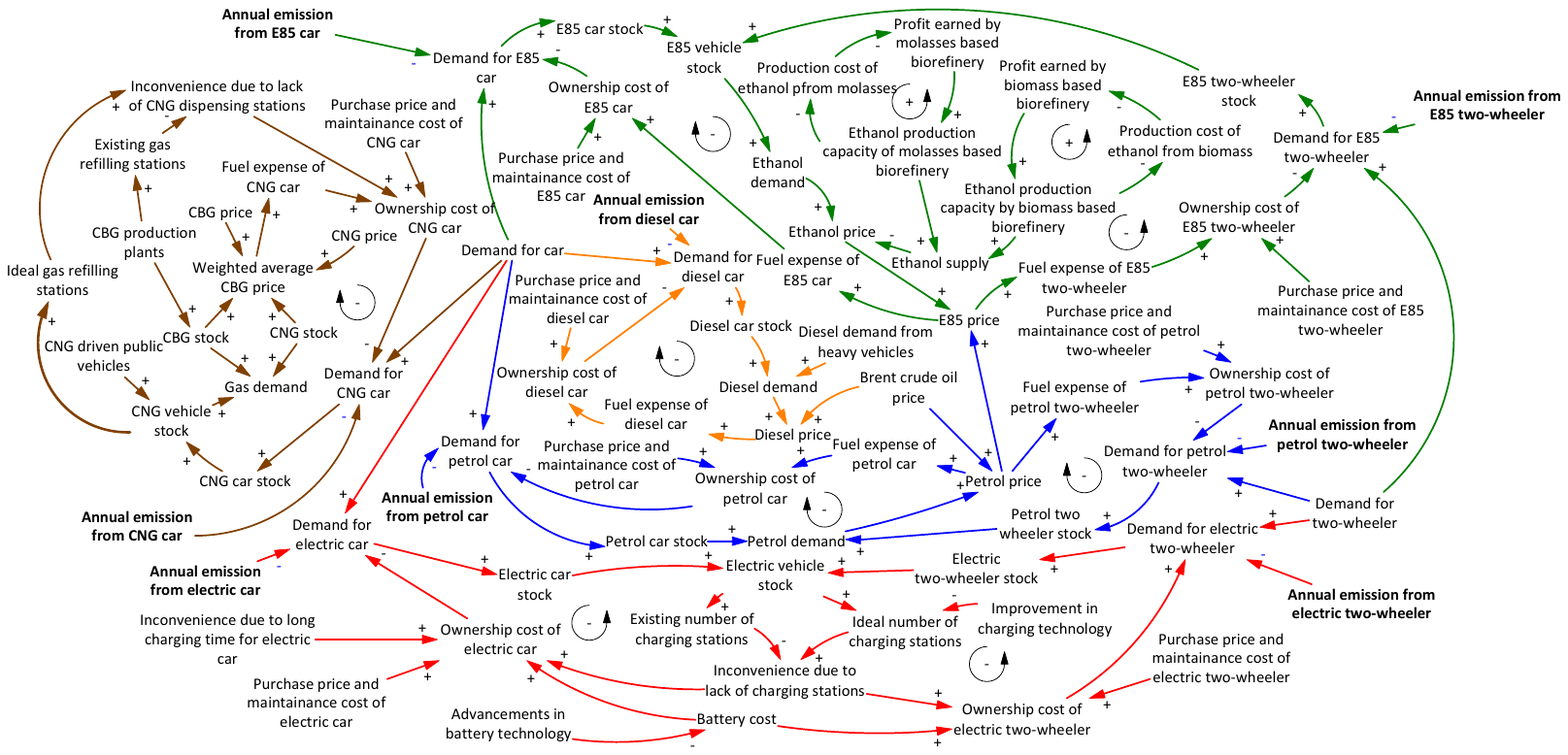}
    \caption{The modified causal loop diagram with the annual emissions from each vehicle option is included as an additional variable. These variables are highlighted in bold in the figure. The ownership cost and annual emissions of each vehicle option negatively affect its annual demand.}
    \label{cld}
\end{figure}

The causal loop diagram developed by \cite{saraf2021development} is modified to include the modifications explained in the preceding sections. The modifications are highlighted in bold in the modified CLD shown in Fig.\ref{cld}. All other variables, their causal relation, and feedback loops of the system are the same as those in \cite{saraf2021development}. Higher annual emissions of the vehicle will reduce its annual demand while lower emissions will increase its demand. Thus, similar to the ownership cost of the vehicles, the annual emissions are also related to the respective annual demands by the negative feedback loop.  

\section{Model Parameterization}
\label{parameterization}

The modified model includes additional parameters that need to be determined in order to perform model simulations. This section describes the estimation of $\mu_p^{Pj}$ and $\mu_e^{Pj}$ values in the model as well as the values of $\alpha^{Pj}_k$. Subsequently, the section presents a discussion of various scenario formulations for the model simulations.

\subsection{Estimation of $\alpha^{Pj}_k$, $\mu_p^{Pj}$ and $\mu_e^{Pj}$}

The basis for parameter estimation was the annual sales data of different options in cars and two-wheeler for the years 2020 and 2021. Additionally, the share for each population subset i.e., $\alpha^{Pj}_k$ for the year 2020 was based on survey data \citep{survey3}. These values are presented in Table \ref{table1}. Values of $\mu_p^{Pj}$ and $\mu_e^{Pj}$ that resulted in a good prediction of the sales values for years 2020 and 2021 were finalized and reported in Table \ref{table1}. A comparison of reported annual vehicle sales (in \%) for 2020 and 2021 with the model output is given in  \ref{comparison} Table \ref{table2}. The comparison shows that the model outputs match well with the reported sales data. Most importantly, the major share of the total vehicle demand, i.e., petrol and diesel, show a good match. Since it is assumed that E85 vehicles sale will start from 2023, the corresponding value is 0\%. The differences in the predicted and actual sales values for EVs and CNG vehicles could be due to multiple reasons. Firstly, the input data ($\alpha^{Pj}_k$) are based on a survey with a small sample size which may not represent the entire population \citep{survey3}. Secondly, annual emissions of electric and CNG vehicles are low (see \ref{ghgperkmcar}). As a result, EVs and CNG vehicles are more preferred than the reported data. Thirdly, the uniform spatial distribution of refueling infrastructure for EVs and CNG is assumed in the model. However, in reality, the distribution of refueling stations is highly non-uniform. This results in higher adoption of EVs and CNG vehicles in the model compared to the actual data.

\begin{table}[t]
\centering
\caption{Estimated values of $\mu_p^{Pj}$ and $\mu_e^{Pj}$ for cars and two-wheelers}
\label{table1}
\begin{tabular}{cccccc}
\hline
\multicolumn{1}{l}{\multirow{2}{*}{Consumer category}} & \multicolumn{1}{l}{\multirow{2}{*}{Share in total population (in \%)}} & \multicolumn{2}{c}{Car}                                   & \multicolumn{2}{c}{Two-wheeler}       \\ \cline{3-6} 
\multicolumn{1}{l}{}                                   & \multicolumn{1}{l}{}                                                  & \multicolumn{1}{l}{$\mu_p$} & \multicolumn{1}{l}{$\mu_e$} & $\mu_p$ & \multicolumn{1}{l}{$\mu_e$} \\ \hline
P1                                                     & 65                                                                    & 430000                      & 10                          & 600000  & 10                          \\
P2                                                     & 20                                                                    & 430000                      & 3000                        & 600000  & 1000                        \\
P3                                                     & 10                                                                    & 10                          & 10                          & 10      & 10                          \\
P4                                                     & 5                                                                     & 10                          & 3000                        & 10      & 1000                        \\ \hline
\end{tabular}
\end{table}

\subsection{Dynamic trends of $\alpha^{Pj}_k$}

As previously mentioned, $\alpha^{Pj}_k$ for each $Pj$ represents the fraction of consumers in a particular decision-making category and this needs to be determined from survey studies. However, the modeling framework developed here can be used to quantify the impact of possible changes in consumer distribution, and thereby also find out the desired distribution for a specific outcome. It is generally expected that an increase in per capita income and greater awareness about climate change will result in the increased importance of emissions in vehicle selection decisions. Thus, $\alpha^{P1}_k$ will reduce with time, while there will be a simultaneous increase in $\alpha^{P2}_k$ and $\alpha^{P4}_k$. The actual realization of this trend is difficult to predict. For simplification, it is assumed here that the reduction in $\alpha^{P1}_k$ value will be equally distributed between $\alpha^{P2}_k$ and $\alpha^{P4}_k$. This assumes that consumers from P1 will become aware of the environmental impacts and shift to either P2 or P4 category. Shifts between other categories are ignored at this stage but can be included with the availability of data. The fraction of consumers in the category in which consumers consider neither ownership cost nor annual emissions in their decision-making is assumed to not change with time. \ref{share} Table \ref{table4} shows that the quantitative distribution of consumers in each category. $x$ represents the awareness rate. Higher awareness rate will rapidly reduce $\alpha^{P1}_k$ and increase $\alpha^{P2}_k$ and $\alpha^{P4}_k$ and vice versa. The changes in $\alpha^{Pj}_k$ with respect to time are not expected to be linear. Instead, it is assumed to follow the Gompertz curve because the adoption of any new technology generally starts very slowly but then increases rapidly and finally saturates at a value. Therefore, $\alpha^{P1j}_k$ is modelled using the Gompertz curve, while $\alpha^{P2}_k$ and $\alpha^{P4}_k$. are modified as explained earlier. 

As mentioned in section \ref{inclusion}, there could be many possibilities on how variations in consumer preferences would evolve in the future. When a maximum fraction of consumers attain environmental awareness, then it is termed as maximum awareness level, after which the $\alpha^{Pj}_k$ will saturate. However, it is difficult to quantitatively estimate the $\alpha^{Pj}_k$ values at saturation due to a lack of data. Therefore, to represent one of the many possibilities for consumer evolution, the saturation values of $\alpha^{Pj}_k$ for P1, P2, and P4 are assumed to be 15, 45, and 30, respectively. This is based on the fact that at the maximum awareness level, the P2 share is expected to be the largest, P4 will be higher than P1, and P3 will be the lowest. This will provide an opportunity to envisage the impacts of a high awareness level on consumers. These saturation values are used to formulate various scenarios described next.

\subsection{Model application}
\label{scenario}

The model was programmed in MATLAB\textsuperscript \textregistered. The simulation horizon was from 2020 to 2050 and the simulation time step was one year. The model has been used to study different aspects related to the transport sector adoption. In the first case, the model studies a set of scenarios capturing the changing priorities of users toward vehicle selection criteria. The second set of scenarios studies policy possibilities/options that may promote the adoption of novel options. The third set of scenarios considers the possibility of a disruptive change in the personal vehicle transport sector. These are the scenarios described next:

\subsection{Changes in decision-making preferences}
\label{scenarios}

The model is first used to study the impact of different awareness trends among consumers and their impact on vehicle adoption as well as GHG emissions. The specific scenarios considered are as follows:

\begin{itemize}
    \item Base scenario - The first scenario, termed as base scenario, assumes that the current status of environmental awareness among consumers will continue in the future. Therefore, the $\alpha^{Pj}_k$ for each consumer category will remain the same as estimated from the survey \citep{survey3}. Although such a scenario is not realistic, it can serve as the basis for comparing the differences and benefits of rising environmental awareness among consumers.
    \item Delayed awareness scenario - The second scenario is termed the delayed awareness scenario as it assumes a rise in environmental awareness among consumers but not in the immediate future but after a few years. Fig. \ref{delayed} (in \ref{description}) shows that the rise in environmental awareness starts around 2030. Maximum awareness reaches around 2040. In the SD model, the annual demand for vehicles is an input to the model. It increases and peaks around 2035 then reduce afterward like an inverted parabola. In this scenario, the rise in environmental awareness coincides with the phase when new vehicle demand is reducing. 
     \item Early awareness scenario - Third scenario, termed as early awareness scenario, assumes a rise of environmental awareness among consumers in the immediate future. Fig. \ref{early} (in \ref{description}) shows that P2 and P4 rise rapidly until the maximum awareness is reached around 2035 unlike in the delayed scenario. Consequently, P1 drops till 2035 before saturating. In this scenario, the rise in environmental awareness coincides with the phase when vehicle demand is also rising rapidly.
\end{itemize}

\subsection{Technology development and policy alternatives} 
\label{technlogical scenarios}

To study the impact of technology development and policy options on vehicle adoption trends, three separate scenarios are formulated as follows:

 \begin{itemize}
     \item Renewable electricity grid: This scenario assumes 50\% penetration of renewable sources in the grid mix by 2030 as per the NDC in COP26. For this, the prediction of the electricity grid mix from 2020 to 2050 was taken from IEA 2021 \citep{electricitygrid}.
     \item Carbon tax: According to International Monetary Fund, the world would need a carbon tax rate of \$ 75 per tonne CO$_2$ (5.7 INR/kg CO$_2$) by 2030 to reduce emissions to a level consistent with a 2$^{\circ}$C warming target \citep{carbontax}. Currently, no explicit carbon tax has been imposed by the Indian government. However, this scenario assumes that an explicit carbon tax will be levied, which will increase linearly from 0 INR/kg CO$_2$ in 2022 to 5.7 INR/kg CO$_2$ by 2030, and thereafter it will remain constant at 5.7 INR/kg CO$_2$ till 2050.
     \item Reduced driving of private vehicles: A report by \cite{vkt} has estimated the reduction in annual vehicle kilometer traveled (vkt) due to the high adoption of ride-sharing and public transits. This scenario assumes a strong development of the public transportation system which leads to the reduction in the annual vkt of private vehicles. The model has considered the annual vkt for cars and two-wheelers from \cite{saraf2021development}. To formulate this scenario, a linear reduction in the annual vkt for cars and two-wheelers is assumed, varying from 14\% of annual vkt in 2020 to 50\% of annual vkt in 2050 as adopted from \cite{vkt}.
     
 \end{itemize}
 
The combination of these scenarios with different awareness scenarios explained previously is also simulated.

\section{Results and discussion}
\label{results}

This section describes the results in two parts. Firstly, the results and related discussions are presented for scenarios considered in section \ref{scenario}. Secondly, the results as well as the implications of the feedback effects of the scenarios mentioned in section \ref{technlogical scenarios} are discussed.

\subsection{Base scenario}

\begin{figure}[!ht]
	\begin{subfigure}{\textwidth}
		\centering
 		\includegraphics[width=15cm,height=15cm,keepaspectratio]{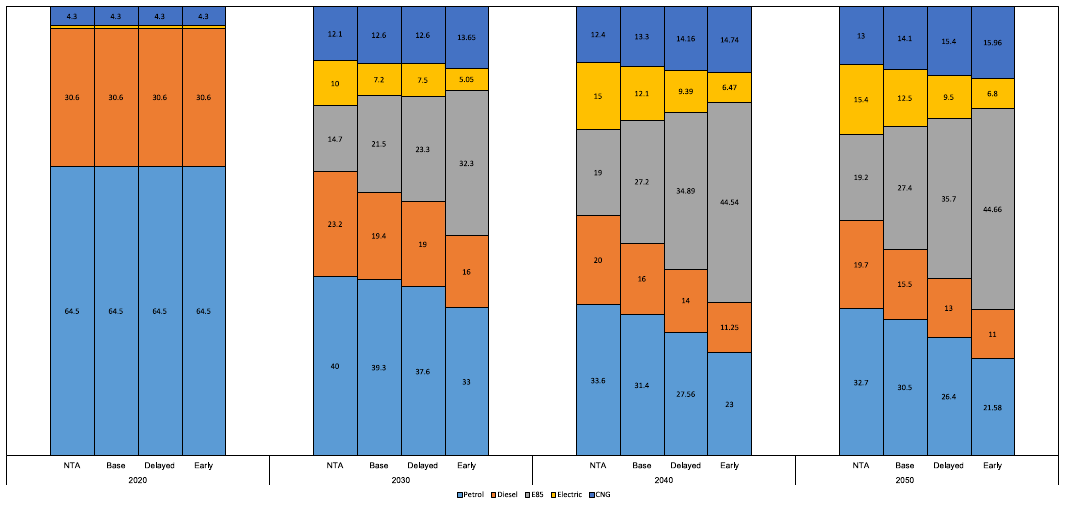}
 		\caption{The adoption trend of various options in cars for the four scenarios. The NTA is the new technology adoption scenario for the previous version of the model. The highlighted numbers on the bars show the share of respective car stock in the total car stock in \%.}
 		\label{car}
	\end{subfigure}
 	\begin{subfigure}{\textwidth}
 		\centering
		\includegraphics[width=15cm,height=15cm,keepaspectratio]{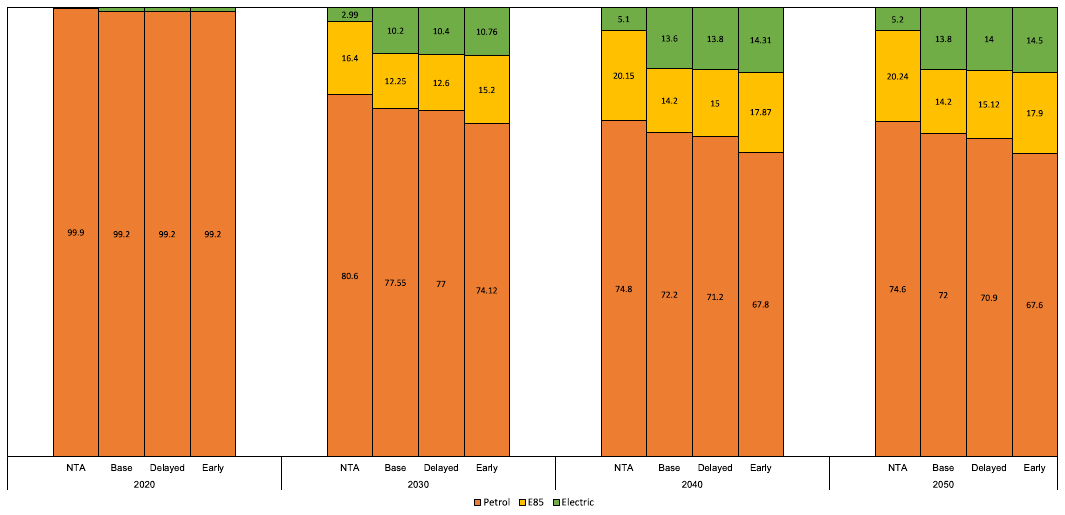}
		\caption{The adoption trend of various options in two-wheelers for the four scenarios. The NTA is the new technology adoption scenario for the previous version of the model. The highlighted numbers on the bars show the share of respective two-wheeler stock in the total two-wheeler stock in \%.}
		\label{twowheeler}
 	\end{subfigure}
 	\caption{Adoption trend of various options in cars and two-wheelers}
 \end{figure}

The simulation results for this scenario show that the car and two-wheeler stocks were comprised majorly of petrol (gasoline) driven vehicles as shown in Fig. \ref{car} and \ref{twowheeler}, respectively. This was expected because, in the base scenario, a large fraction of consumers preferred low-cost options. With regards to the car stock, a reduction of about 40\% and 22\% was observed in the shares of petrol and diesel car stocks, respectively from 2020 to 2030. This was due to the adoption of other vehicle options, i.e., E85, EV, and CNG vehicles. The plots of GHG emission per km for all various options in the car and two-wheeler are shown in Fig. \ref{ghgperkmcar} and \ref{ghgperkmtw}, respectively in Appendix \ref{modelinputs}. Among the novel vehicle options, the E85 car had the highest share because of the comparable ownership cost to a petrol car and the least annual emissions of all options. Although the annual emissions of CNG and electric cars were lower than those of petrol and diesel cars, their minor shares in total car stock were due to higher ownership costs owing to a lack of sufficient refueling stations. With regards to two-wheeler stock, the annual emissions for E85 and electric two-wheeler were significantly lower than that of petrol (see Fig. \ref{ghgperkmtw} in \ref{modelinputs}). However, it wasn't sufficient to ensure a high adoption rate of these options among two-wheelers, which were majorly petrol-dominated. Interestingly, the E85 and electric two-wheelers had a comparable share in the total two-wheeler stock. This was unlike in car stock, where E85 share was more than double the share of electric. The electric two-wheeler adoption improved after 2030 because of the reduction in ownership cost. This came from the assumption that by 2030 electric vehicles will have parity with petrol vehicles in terms of purchase prices \citep{ev}. An electric two-wheeler requires less charging time than an electric car, and therefore lack of sufficient charging stations does not affect the ownership cost of the EV two-wheeler as much as an electric car. Therefore, the adoption of the electric car was not much affected by the reduction in the purchase price. EV cars and two-wheeler sales reach 25\% share in the total sales of new vehicles for the year 2030, which is less than the target value of 30\% as per the EV30@30 scheme. As per the CRISIL research group, the EV share in car and two-wheeler stock in 2030 were 4-5\% and 11-13\%, respectively \citep{crisil}. This matches with the simulated EV share of 8\% and 10\% for car and two-wheeler stock for 2030. The ethanol blend fraction in petrol for this scenario reached 13.87\% in 2025, which is again lower than the target value of 20\% as per the National Biofuel policy. However, the predicted ethanol blending rate is close to the forecasted value of 15\% by 2025 by the CRISIL research group \citep{crisil}. Additionally, the blend fractions for the years 2020, 2021 and 2022 were 3.1\%, 6.9\% and 9.3\%, respectively which is close to the actual values of 5\%, 8.1\%, and 10\%, respectively \citep{8.1blend}.

It is important to understand the impact of modeling consumer diversity on the results. For that, results of cars and two-wheelers adoption and GHG emissions were compared with the NTA (New technology adoption) scenario presented by \cite{saraf2021development}. In that case, only annual cost was used as a decision-making factor and only the P1 category of consumers was considered. The comparison of cars and two-wheelers for the two scenarios is shown in Fig. \ref{car} and \ref{twowheeler}, respectively. In the NTA scenario, the preference for options in cars in decreasing order was petrol, diesel, E85, electric, and CNG. However, when environmental awareness was considered in the purchase decision, the preference was changed to petrol, E85, diesel, CNG, and electric. The cumulative adoption of novel vehicle options, i.e., E85, EV, and CNG was 47.5\% of the total private vehicle stock by 2050 in the NTA scenario, whereas the share increased to 54\% in the base scenario. The petrol demand coming from car and two-wheeler stock reduced by 4.4 billion litres by 2050 due to the reduction in petrol vehicle adoption. Similarly, the diesel demand from the car stock also reduced by 7 billion litres by 2050. The ethanol blending rate in 2025 increased from 8.18\% in the NTA scenario to 13.87\% due to increased adoption of E85 vehicle stock. With regards to two-wheelers (see Fig. \ref{twowheeler}), the preference was observed to remain unchanged. The E85 and electric two-wheeler stock were increased while the petrol two-wheeler stock was reduced.  

\begin{figure}[!ht]
    \centering
    \includegraphics[width=15cm,height=15cm, keepaspectratio]{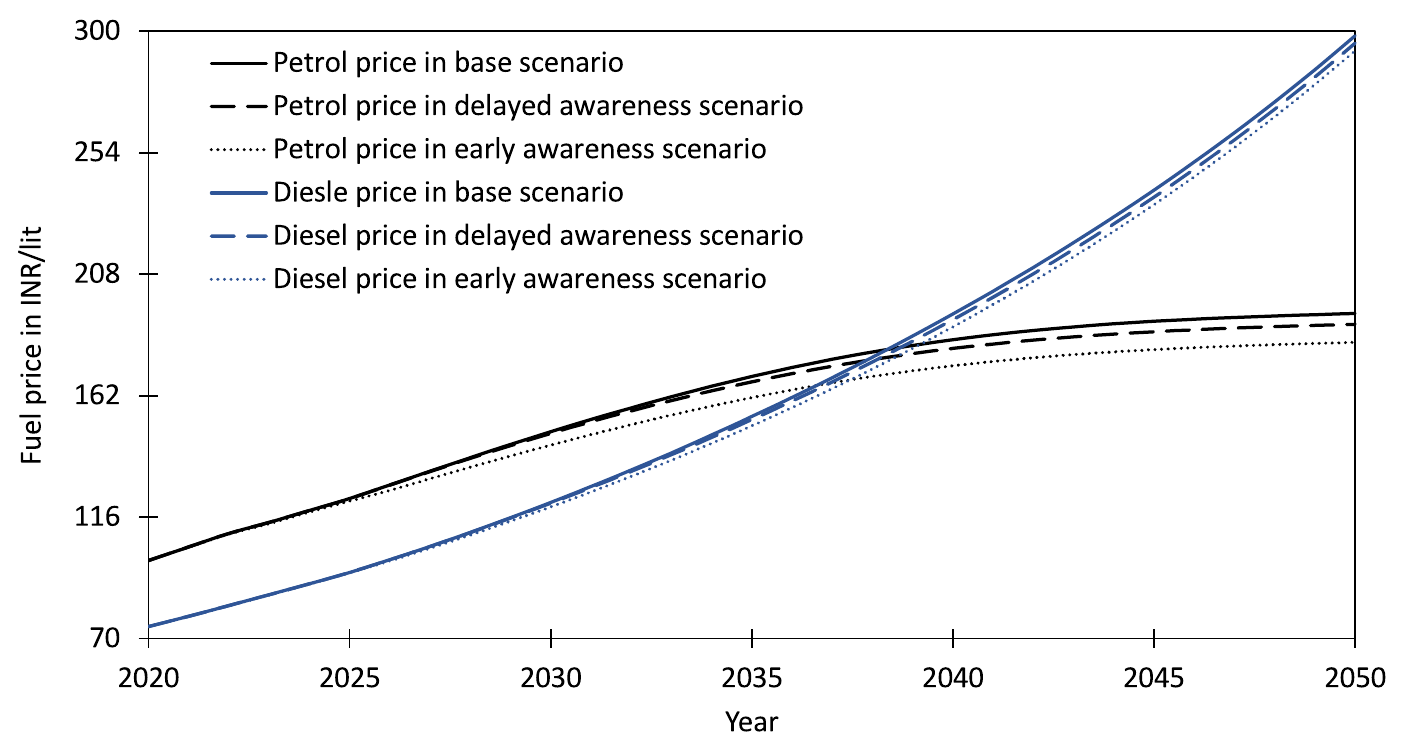}
    \caption{This figure shows the petrol and diesel prices for the three scenarios. A feedback effect of an increase in awareness causes a reduction in the adoption of petrol and diesel vehicle, thereby reducing the respective fuel demand as well as respective fuel prices.}
    \label{fuelprices}
\end{figure}

Fig. \ref{fuelprices} shows the petrol and diesel prices in INR/lit for this scenario. The petrol car and two-wheeler shares saturate at 30\% and 72\%, respectively between 2040 to 2050 due to saturation in vehicle demand. Due to this the petrol demand also saturated, resulting in the stabilization of petrol price at 193 INR/lit (8.77 \$/gal) in 2050 as seen in Fig. \ref{fuelprices}. However, the diesel price rises throughout due to an increase in diesel demand coming from other sectors, i.e., freight transport, power, and agriculture. The diesel price in 2050 was 298 INR/lit (13.54 \$/gal).

\begin{figure}[!ht]
    \centering
    \includegraphics[width=15cm,height=15cm, keepaspectratio]{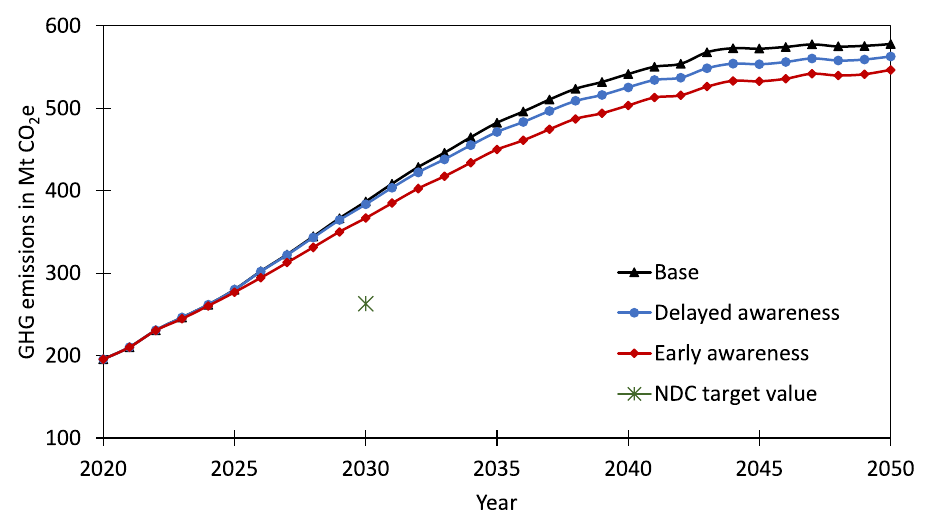}
    \caption{This figure shows the GHG emission trend for the three scenarios and the COP26 emission reduction target value for the road transport sector of India.}
    \label{ghg}
\end{figure}

The GHG emission from private vehicle stock for this scenario is shown in Fig. \ref{ghg}. As the vehicle stock saturates, the GHG emissions also tend to saturate. As per the Paris Agreement target, India's GHG emission should be within 3.8-3.9 billion tonnes by 2030 \citep{saraf2021development}. India has pledged a higher target value of 45\% reduction in emission intensity of 2005 level by 2030 at the COP26 summit. On calculating the target value compatible with COP26 emission reduction, the GHG emission is 3.02 billion tonnes by 2030. Assuming that 10\% of total GHG emissions would be caused by the transport sector and 87\% would be due to the road transport sector \citep{singh2019greenhouse}, the GHG emission target for the road transport sector should be 263 million tonnes by 2030. However, the GHG emission value was 392 million tonnes CO$_2$e in 2030, which was almost 50\% higher than the COP26 target value. This shows that with present awareness among consumers, it is inefficient to meet the GHG emission target.

\subsection{Delayed awareness scenario}
\label{delay}

Compared to the base scenario, this scenario showed a very minor change in the adoption trend and target values. A comparison of shares of various options in cars and two-wheelers between the base and delayed awareness scenario is shown in Fig. \ref{car} and \ref{twowheeler}, respectively. After 2025, the share of the P2 category is the highest followed by P4 then P1. This means that low ownership cost with low emissions options was most preferred. This results in a peculiar observation that as the awareness rate increases throughout the simulation horizon, the electric car share reduces, while the electric two-wheeler share increases (see Fig. \ref{car} and \ref{twowheeler}). This is because options like E85 and CNG cars have lower ownership costs and annual emissions than electric cars. Thus, they become a more attractive option to consider over the electric car. As P2 share increases, E85 and CNG car options dominate over the electric car. However, this is not the case with two-wheeler stock, because the available options are very limited, hence the E85 and electric share increases while petrol decreases. As a result, the fraction of electric vehicle sales in the new sales was slightly reduced from 25\% in base to 24\% in this scenario. In addition to that, the ethanol blend was 13.88\%, almost the same as that in the base scenario.

Petrol and diesel prices for this scenario are shown in Fig. \ref{fuelprices}. This depicts the feedback effect of the system. Due to increased awareness, the petrol and diesel options were less preferred. Consequently, petrol and diesel demand were reduced compared to the base scenario and therefore, the fuel prices were slightly lower. The petrol and diesel prices were 189 INR/lit (8.58 \$/gal) and 295 INR/lit (13.4  \$/gal), respectively in 2050.

The GHG emission trend for the delayed awareness scenario is shown in Fig. \ref{ghg}. As the adoption trend was nearly similar to the base scenario, there was a marginal difference in the GHG emission trends between both scenarios. The GHG emission in 2030 was 388 million tonnes CO$_2$e, higher than the target value of 263 million tonnes CO$_2$e. These results show that an increase in awareness later in the time horizon has little impact on the adoption of new vehicle options as well as the GHG emissions. 

\subsection{Early awareness scenario}

Here the P2 subset rises from the start of the simulation, therefore a more significant reduction in petrol and diesel share was observed compared to the base scenario (see Fig. \ref{car}). As more people preferred low-cost low emission options from the start, electric car adoption was further reduced compared to the delayed scenario. This was taken over by E85 and CNG car shares. Among two-wheelers, the petrol share reduced while E85 and electric share increased as expected from the trend (see Fig. \ref{twowheeler}). The overall share of electric vehicle sales in the new vehicle sales was found to be 21.76\% in 2030, which was lower than the target value.

The ethanol blend was 14.05\% in 2025, which was still lower than the 20\% target. Compared to the delayed awareness scenario (Fig. \ref{fuelprices}), fuel prices were considerably reduced due to reduced adoption of petrol and diesel prices. The petrol price was stabilized at 182 INR/lit (8.26 \$/gal) and the diesel price was at 292 INR/lit (13.26 \$/gal) in 2050. The GHG emission (Fig. \ref{ghg}) was estimated as 371 million tonnes CO$_2$e, which was the lowest among the three scenarios but still  40\% higher than the target value.

An important conclusion from these results is the impact of timing of increased awareness. Increased environmental awareness during the phase when vehicle demand is expected to rise rapidly is critical to make an impact. On comparing the delayed and early awareness scenarios, low emission options, i.e., E85, electric, and CNG vehicle stock were higher in the early awareness scenario even though both scenarios reached the same awareness level at the end. Thus, it is important to encourage the adoption of low-emission options when vehicle demand is also rising.

\subsection{Impact of considering only grid-related emissions for EVs}

One of the unexpected results from the previous studies was the reduction in the adoption of EVs with an early increase in environmental awareness. 

\begin{figure}[!ht]
    \centering
    \includegraphics[width=15cm,height=15cm, keepaspectratio]{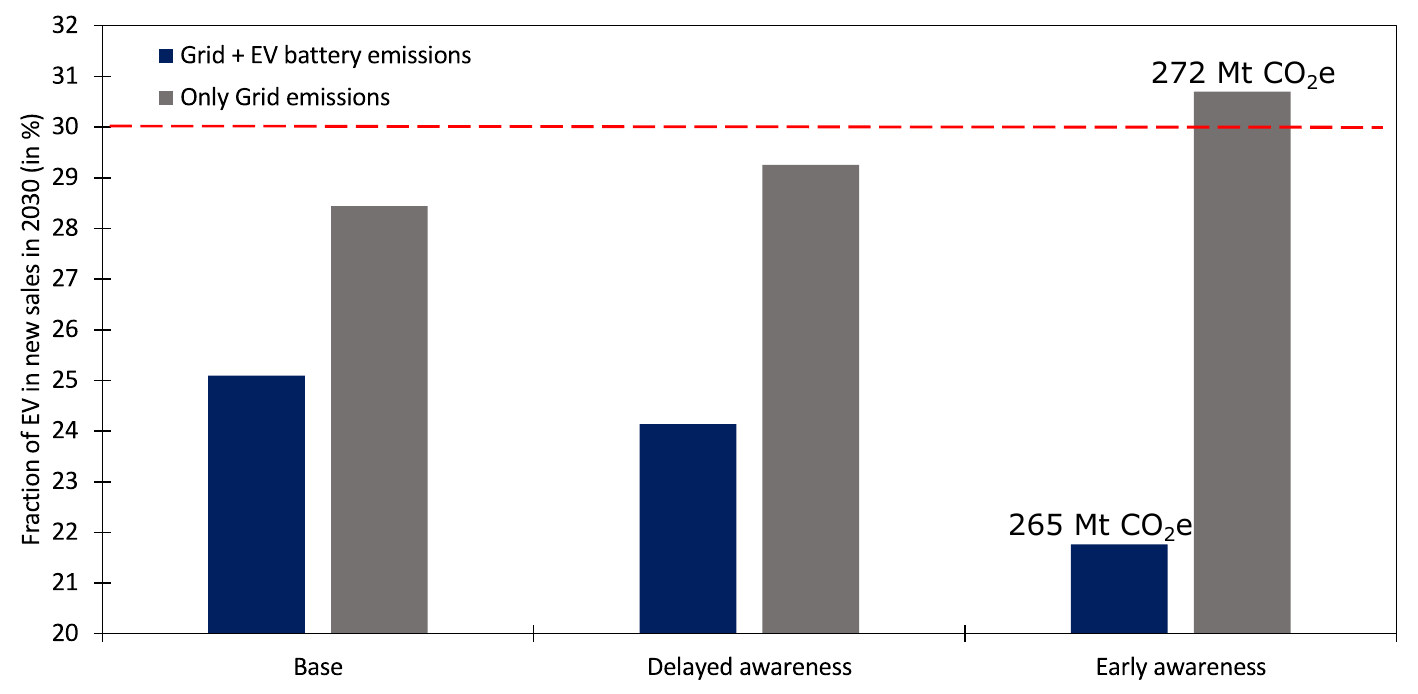}
    \caption{Impact of neglecting the EV battery emissions in the purchase decision. The numbers mentioned above the bars are the GHG emissions in 2030. The adoption will increase with awareness on ignoring battery emissions. The red dashed line in the plot shows the 30\% target value of EV sales as per the EV30@30 policy.}
    \label{batteryemissions}
\end{figure}

As mentioned in section \ref{cld}, the annual emission for EVs is the sum of emissions from the electricity grid used for charging and emissions during the EV battery manufacturing, recycling, and discarding phase. These emissions related to the battery are annualized considering the life of the battery. The inclusion of battery-related emissions led to EVs having higher GHG emissions per km driven than E85 and CNG vehicles, and that caused a reduction in EV adoption. It can though be argued that consumers rarely think from a life cycle perspective and the EVs are perceived to be low-emission vehicles due to potentially low emissions related to electricity used in charging. To test this possibility, a scenario was simulated in which the annual emission of EVs included only the grid emissions. The results for the base, delayed, and early awareness scenarios considering only the grid emissions (Fig. \ref{batteryemissions}) showed that EV adoption increased when only grid emissions were considered. The ownership cost of EV was the same for both scenarios. The EV sales increased from 21.76\% to 30.7\%, meeting the EV30@30 target value for the early awareness scenario. However, this results in increased actual emissions. The actual GHG emissions in 2030 were 380 Mt CO$_2$e, higher than 371 Mt CO$_2$e when consumers were assumed to be aware of battery-related emissions as well.
Even though the difference in GHG emissions for one year is not significant but its cumulative addition for subsequent years might be concerning. Additionally, the model only considers the climate change impact, but other impacts related to water and soil emissions due to EV batteries were not even considered. This result highlights the importance of considering batteries in decision-making. Currently, local battery manufacturing, as well as recycling systems in India are almost nonexistent. Most of the batteries are imported and battery recycling systems are not established, making this a critical aspect of planning. 

%\subsection{Impact of fast development of charging infrastructure}

%The early awareness scenario was simulated by assuming 10 times (i.e., 1 charging station per 100 electric vehicles) faster setting of charging infrastructure than the base scenario (i.e., 1 charging station per 1000 electric vehicles). Faster installation of charging infrastructure will reduce the inconvenience associated with refueling EVs. Thus, this reduced the inconvenience cost as well as the ownership cost of EVs. The result showed that electric vehicle sales share in the new vehicle sales was 32\% in 2030, higher than the target value as per the EV30@30 scheme. Thus, the combination of early awareness and fast setting of supporting infrastructure can help in meeting the target value.  

\subsection{Impact of technology development and policy option}

The results for combinations of technology development and policy option scenarios with environmental awareness scenarios (see section \ref{scenarios}) are presented in Table \ref{scenarioresults}. It can be seen that in all cases, the desired targets were not achieved. Apart from the EV30@30 target, the ethanol blending of 16.3\% and the GHG emissions of 272 Mt CO$_2$e were the closest to the target values achievable. The results of these scenarios are discussed next.

\begin{itemize}
    \item Renewable electricity grid with environmental awareness scenarios: As per the causal relationships (Fig. \ref{feedback1}), the increase in renewable share in the electricity mix reduces the grid emissions as well as electricity price \citep{electricityprice}. Lower emissions will increase the adoption driven by environmental awareness. Additionally, lower electricity price will reduce the fuel expense, thereby reducing the ownership cost of EVs. Both these factors promote the adoption of EVs (Table. \ref{scenarioresults}). The ethanol blending is calculated as the amount of ethanol produced for blending per litre of petrol consumed. As EV adoption increases due to the renewable grid, the adoption of petrol vehicles reduces, reducing petrol consumption. Therefore, the ethanol blend fraction increased from base to early awareness scenarios for a greater share of renewable in the electricity grid. 
    
    \item Carbon tax with the environmental awareness scenarios: Imposition of carbon tax penalizes the high emission options such as petrol and diesel. Based on the causal relationships (Fig. \ref{feedback4}) adoption of E85 vehicles increases due to low emissions and low carbon tax compared to other vehicles. This increases the E85 stock, thereby increasing the ethanol demand for blending. Petrol demand will reduce due to the reduced adoption of petrol vehicles. This will increase the ethanol blending rate as shown in Table. \ref{scenarioresults}. 
    
    \item Reduced private vehicle driving with the environmental awareness scenarios: When renewable electricity grid is combined with reduced vehicle kilometer travelling, it leads to an interesting feedback effect of the system. Fig. \ref{feedback3} shows that an increase in renewables share increases the EV adoption via environmental awareness and ownership cost of the vehicle. Consequently, this results in reduced adoption of other vehicle options including petrol vehicles. Consequently, the vehicle stock reduces, thereby reducing the petrol demand. This lowers the petrol price as shown by the positive sign in the figure. If the vehicle travelling is also reduced, it further reduces the petrol demand as well as its price. This results in low fuel expense of the vehicle, resulting in higher demand for petrol vehicles due to lower ownership cost. The adoption of petrol vehicles dominates over the adoption of EVs, therefore a significant reduction in EV adoption can be observed in Table. \ref{scenarioresults}. 
\end{itemize}

With renewable electricity grid or carbon tax, there were some feedback effects to support the adoption of EVs. If reduced vehicle travelling is alone simulated, it results in very low adoption of EVs (see Table. \ref{scenarioresults}). Due to reduced petrol consumption, the ethanol blend is very high. As reduced vehicle driving was due to improved public transportation. It is important to note that GHG emissions are considerably low in the private road transport sector, however, it would have increased in the public road transport sector.

\begin{landscape}
\begin{table}[]
\centering
\caption{Comparison of various scenario combinations in terms of government targets regarding GHG emissions, ethanol blending, and EV adoption. Those values which meet the target value are highlighted in bold}
\label{scenarioresults}
\resizebox{\columnwidth}{!}{%
\begin{tabular}{cccccccccc}
\hline
\multirow{2}{*}{Target values for comparison} &
  \multicolumn{3}{c}{COP26 target - 263 Mt CO$_2$e} &
  \multicolumn{3}{c}{Target value - 20\%} &
  \multicolumn{3}{c}{EV30@30 - 30\% sales by 2030} \\
 &
  \multicolumn{3}{c}{GHG emission in 2030 (in Mt CO$_2$e)} &
  \multicolumn{3}{c}{Ethanol blending in 2025 (in \%)} &
  \multicolumn{3}{c}{\% of EV in new sales} \\
Scenario combinations              & Base   & Delayed awareness & Early awareness & Base   & Delayed awareness & Early awareness & Base   & Delayed awareness & Early awareness\\ \hline
Electricity grid                   & 378.7 & 375.4     & 358.7   & 13.89 & 13.9   & 14.08 & 29.51 & \textbf{30.51}    & \textbf{32.7} \\
Carbon tax                         & 387.8 & 384.7  & 369.4 & 13.89  & 13.89   & 14.07  & 26.69 & 25.45   & 22.38 \\
Reduced driving                    & 299.2 & 296.25  & 281.71 & 16.03  & 16.03   & 16.27 & 16  & 15.62    & 14.64  \\ \hline
Electricity grid + Carbon tax      & 374.7 & 371.8   & 356.5 & 13.91 & 13.92   & 14.09 & \textbf{31.77} & \textbf{32.37}   & \textbf{33.65} \\
Electricity grid + reduced driving & 291.5 & 288.65   & 273.9 & 16.05  & 16.06   & 16.3 & 20.88 & 22.7      & 26.83  \\
Carbon tax + reduced driving       & 295.9 & 293.2   & 279.8   & 16.05 & 16.06   & 16.3  & 18.4 & 17.6    & 15.69 \\ \hline
All scenarios together             & 288.2 & 285.6   & 272.1   & 16.08 & 16.08    & 16.31 & 23.83 & 25.2    & 28.29 \\
None  of the scenario              & 391.72   & 388.3   & 371.72 & 13.87 & 13.88   & 14.05  & 25.09 & 24.14    & 21.76 \\ \hline
\end{tabular}%
}

\end{table}
\end{landscape}

\section{Conclusion}
\label{conclusion}

The objective of this work was to understand the adoption trend of low-emissions vehicle options i.e., E85, electric, and CNG for the private road transport sector of India as environmental awareness among consumers evolves with time. The system dynamics model developed earlier was extended to include environmental awareness of consumers impacting their purchase decision. Hence, the purchase decision was formulated as a function of ownership cost and annual emissions of the vehicle.  
A methodology was developed to capture the variations in the adoption trend as environmental awareness evolves among consumers. The simulation results showed that considering environmental awareness in the purchase decision increased the adoption of low-emission options, i.e., E85, EV, and CNG. Low-emission vehicle options constituted 54\% of the total vehicle stock in 2050. However, the impact of the timing of increased environmental awareness is more important here. Increased environmental awareness during the phase of rapidly rising vehicle demand makes a critical impact. In addition to that, accounting for both grid as well as battery emissions in the purchase decision is important as battery emissions are very significant. Government targets regarding ethanol blending, EV adoption, and GHG emission reduction are very optimistic, even technology development and various policy options such as carbon tax and increased public transportation are insufficient to meet the targets.

Some limitations of this work are summarised as follows. Firstly, attributes such as brand loyalty, societal status, and vehicle performance such as acceleration, fuel economy, etc. are not considered in the purchase decision. Secondly, for simplification, broad categories of consumers were considered in the model. However, there might be a complex structure of categories among consumers that will be very difficult to predict and model. Thirdly, the model results are the predictions of one of the many future possibilities. Additionally, the model was parameterized based on recent years' data. Therefore, the model might be required to be re-parameterized as more data become available. Lastly, the transport sector of India could undergo several changes, some of which could be disruptive such as the adoption of hydrogen fuel cell vehicles. Therefore, the model is required to be updated in order to address these changes in the future.    

This work can be extended to evaluate the impact of the adoption of E85, EV, and CNG vehicles on other environmental impact categories such as acidification potential, particulate matter, photochemical oxidation, etc. This would be particularly useful for estimating impacts due to EV adoption. Change in adoption trends due to deteriorating vehicle mileage and scrapping policy can be analyzed using this model to estimate more accurate fuel demand and related emissions.

\section*{Declarations}
%\label{declr}
%\begin{enumerate}
%	\item \textbf{Funding}: This study was funded by the Department of Biotechnology (DBT), Government of India, through grant BT/EB/PAN IIT/2012.
\textbf{Funding}: The authors wish to acknowledge support by SPLICE - Climate Change Programme, Department of Science and Technology, Ministry of Science and Technology through project DST/CCP/CoE/140/2018 (G).

\newpage
\bibliographystyle{elsarticle-harv} 
\bibliography{cas-refs}

\newpage

\appendix
\appendixpage

\section{Comparison of reported vehicle sales data with the simulation results obtained for 2020 and 2021}
\label{comparison}

\begin{table}[h]
\centering
\caption{Comparison of reported vehicle sales data with the simulation results obtained for the year 2020 and 2021}
\label{table2}
\resizebox{\columnwidth}{!}{%
\begin{tabular}{ccccc}
\hline
Vehicle options &
  \multicolumn{1}{l}{Reported sales in 2020} &
  \multicolumn{1}{l}{Model output for 2020} &
  \multicolumn{1}{l}{Reported sales in 2021} &
  \multicolumn{1}{l}{Model output for 2021} \\ \hline
Petrol car                          & 75\%   & 69\% & 74.1\% & 62\%    \\
Diesel car                          & 18.7\% & 17\% & 18.4\% & 20.6\%  \\
E85 car                             & 0\%    & 0\%  & 0\%    & 0\%     \\
Electric car                        & 0.2\%  & 5\%  & 0.4\%  & 6.87\%  \\
CNG car                             & 6.1\%  & 9\%  & 7.1\%  & 10.53\% \\
Petrol two-wheeler                  & 99\%   & 91\% & NA     & NA      \\
\multicolumn{1}{l}{E85 two-wheeler} & 0\%    & 0\%  & NA     & NA      \\
Electric two-wheeler                & 1\%    & 9\%  & NA     & NA      \\ \hline
\end{tabular}%
}
\end{table}

\section{Share distribution among different consumer categories}
\label{share}

\begin{table}[h]
\centering
\caption{Share distribution among different consumer categories}
\label{table4}
\begin{tabular}{ccc}
\hline
\multicolumn{1}{l}{Consumer category} & \multicolumn{1}{l}{Share in total consumer population for the year 2020 (in \%)} &    \multicolumn{1}{l} {Variation in share}              \\ \hline
P1                                    & 65                                                   & 65-$x$           \\
P2                                    & 20                                                   & 20+$\frac{x}{2}$ \\
P3                                    & 10                                                   & 10               \\
P4                                    & 5                                                    & 10+$\frac{x}{2}$ \\ \hline
\end{tabular}
\end{table}

\section{Description of delayed and early awareness scenarios}
\label{description}

\begin{figure}[!ht]
	\begin{subfigure}{\textwidth}
		\centering
 		\includegraphics[width=13cm,height=13cm,keepaspectratio]{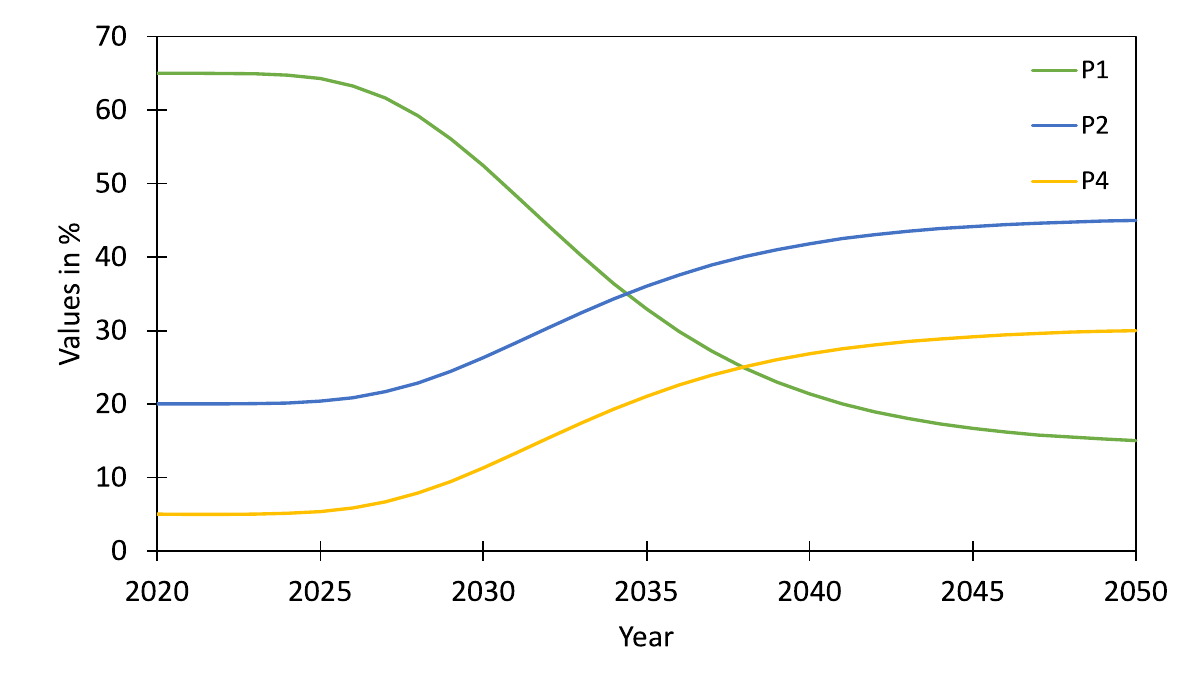}
 		\caption{The change in P1, P2, and P4 values with time for the delayed awareness scenario}
 		\label{delayed}
	\end{subfigure}
 	\begin{subfigure}{\textwidth}
 		\centering
		\includegraphics[width=13cm,height=13cm,keepaspectratio]{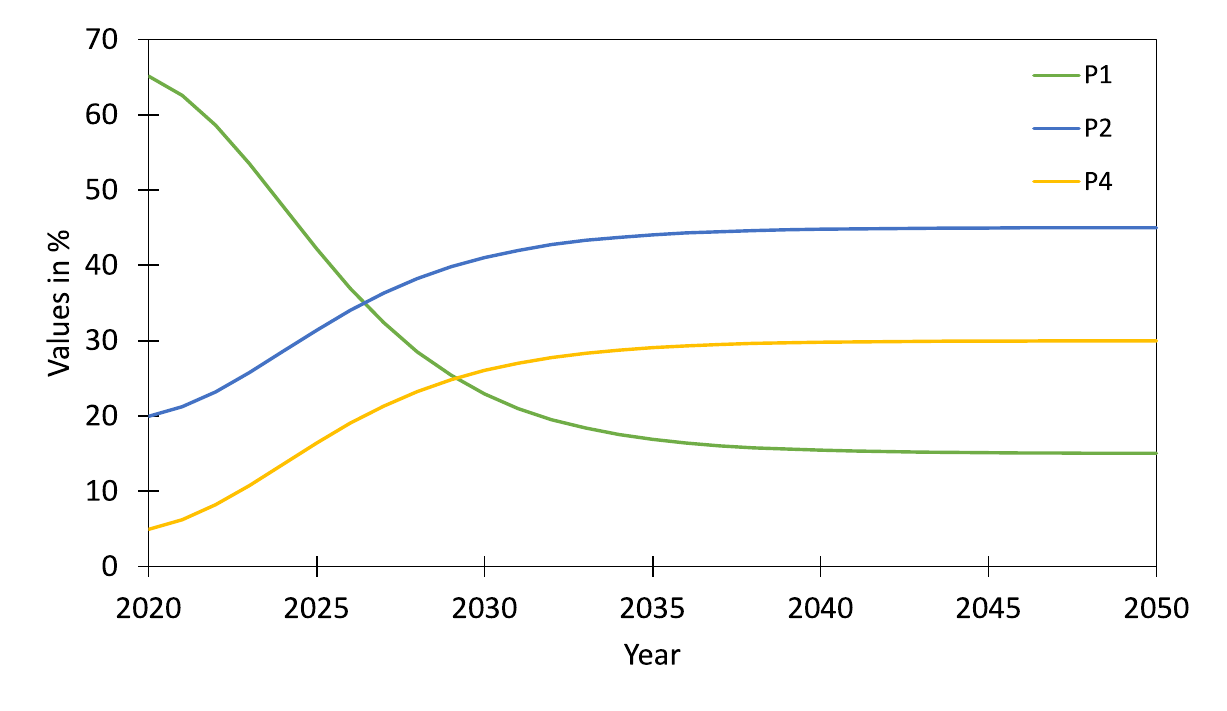}
		\caption{The change in P1, P2, and P4 values with time for the early awareness scenario}
		\label{early}
 	\end{subfigure}
 	\caption{Description of delayed and early awareness scenarios}
 \end{figure}

\section{GHG emissions and ownership cost trends for various vehicle options}
\label{modelinputs}

\begin{figure}[!ht]
    \centering
    \includegraphics[width=17cm,height=17cm, keepaspectratio]{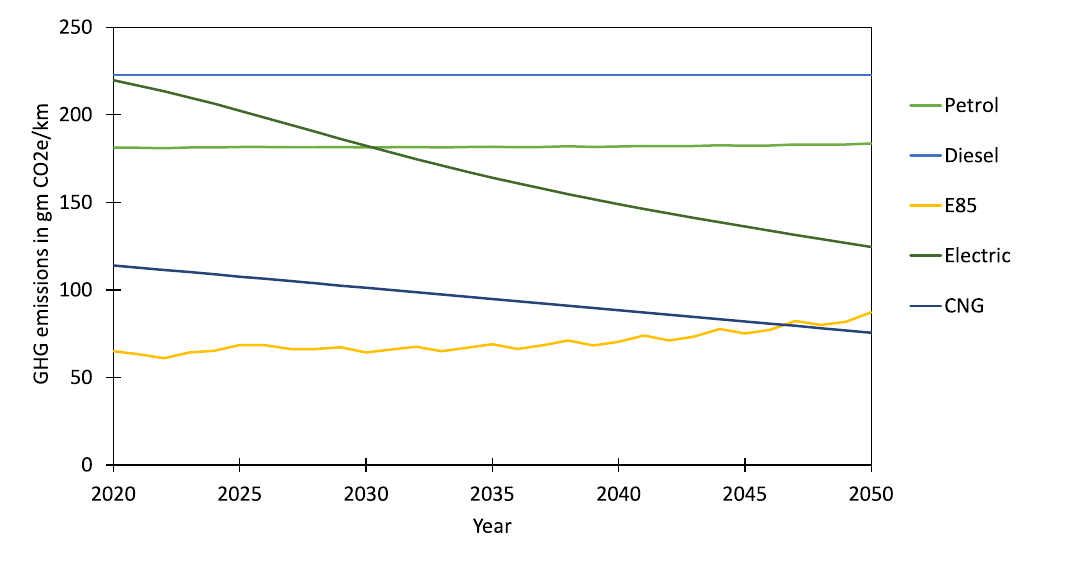}
    \caption{This figure shows the GHG emission in gm CO$_2$e per km for various options in cars.}
    \label{ghgperkmcar}
\end{figure}

\begin{figure}[!ht]
    \centering
    \includegraphics[width=17cm,height=17cm, keepaspectratio]{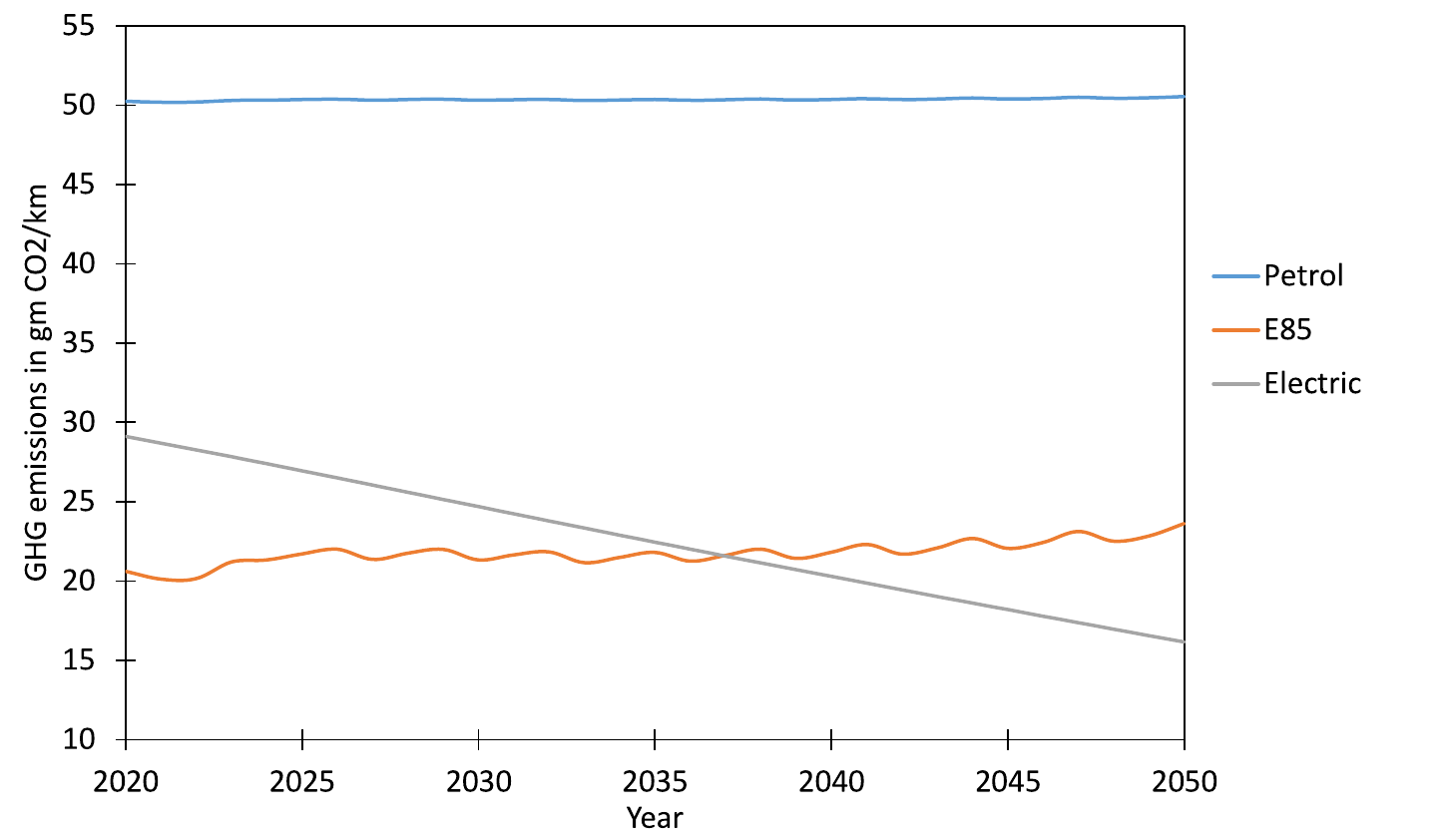}
    \caption{This figure shows the GHG emission in gm CO$_2$e per km for various options in two-wheelers.}
    \label{ghgperkmtw}
\end{figure}

The plots for GHG emission per km for various options in car and two-wheeler are shown in Fig. \ref{ghgperkmcar} and \ref{ghgperkmtw}. The GHG emission for E85 vehicle is the  weighted average emission of ethanol obtained from various feedstocks, i.e., molasses, sugar, damaged food grains, and biomass. The share of molasses and biomass based ethanol fluctuates due to profit/loss experienced by the respective biorefineries. As a results a slight fluctuation is observed in the GHG emission per km trend for E85 vehicles. 

\begin{figure}[!ht]
    \centering
    \includegraphics[width=17cm,height=17cm, keepaspectratio]{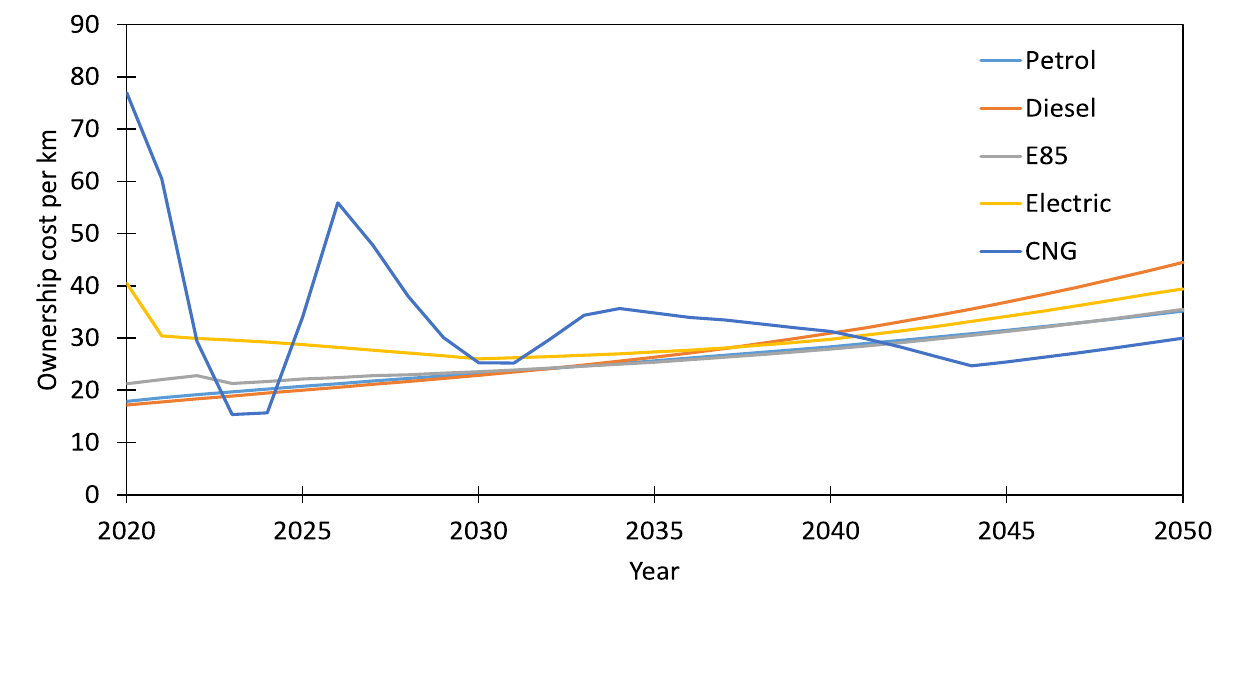}
    \caption{This figure shows the ownership cost in INR per km for various options in car.}
    \label{costperkm}
\end{figure}

The plot of ownership costs of various car options is shown in Fig \ref{costperkm}. The ownership cost of CNG car is the sum of purchase price, fuel expense, maintenance cost, and inconvenience cost due to insufficient gas refilling stations. The inconvenience cost forms about 50\% of the ownership cost of the vehicle. Thus, the adoption of CNG vehicle is very sensitive to the availability of gas refilling stations. Fluctuations in the ownership cost of CNG vehicle arises due to a negative feedback loop. Adoption of CNG vehicles increases the demand for CNG stations. Lack of sufficient CNG stations increases the inconvenience cost of CNG vehicles, thereby increasing the ownership cost of the vehicle. As a result, the adoption reduces making the existing number of CNG vehicles sufficient for the stock. Hence the loop continues resulting in fluctuating ownership cost of CNG vehicle  

\section{Causal loop diagrams showing various combination of scenarios}
\label{feedback}

\begin{figure}[!ht]
    \centering
    \includegraphics[width=17cm,height=17cm, keepaspectratio]{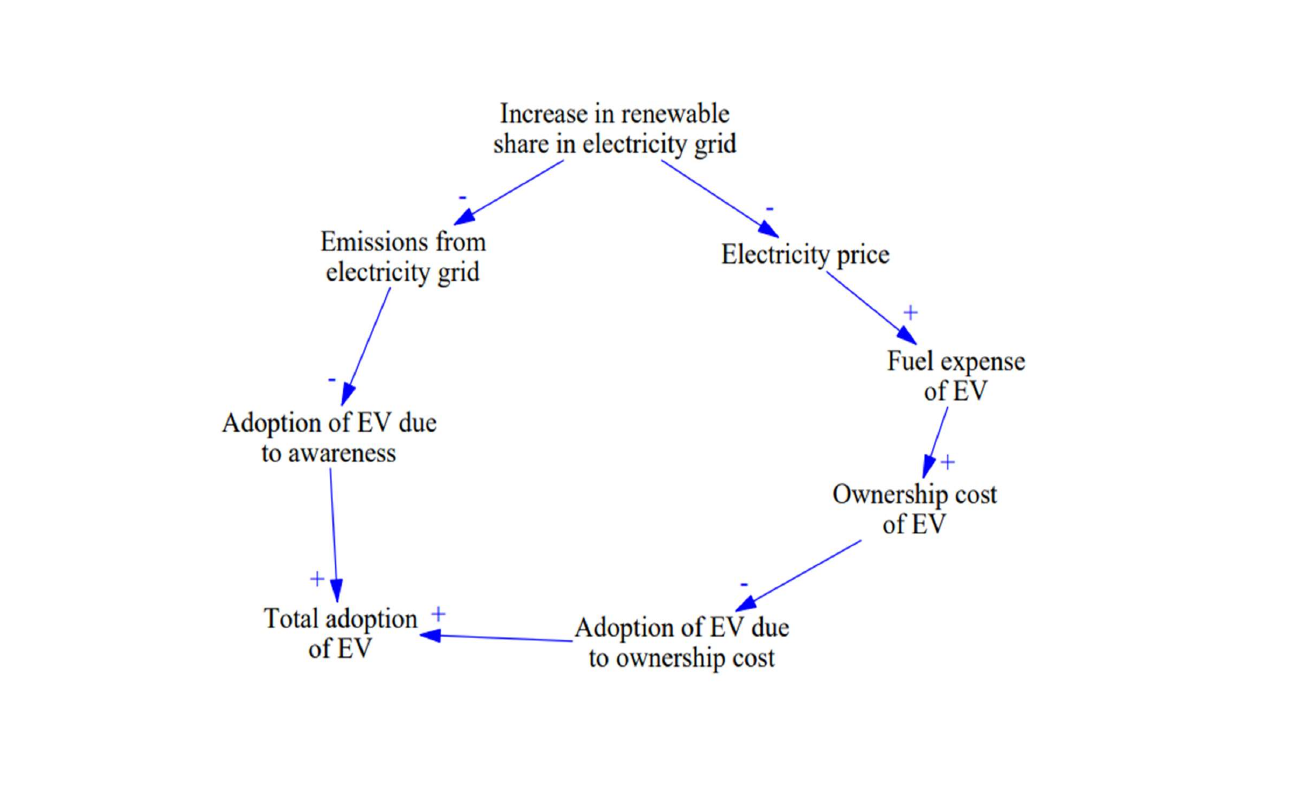}
    \caption{Impact of greener electricity grid on EV adoption}
    \label{feedback1}
\end{figure}

\begin{figure}[!ht]
    \centering
    \includegraphics[width=17cm,height=17cm, keepaspectratio]{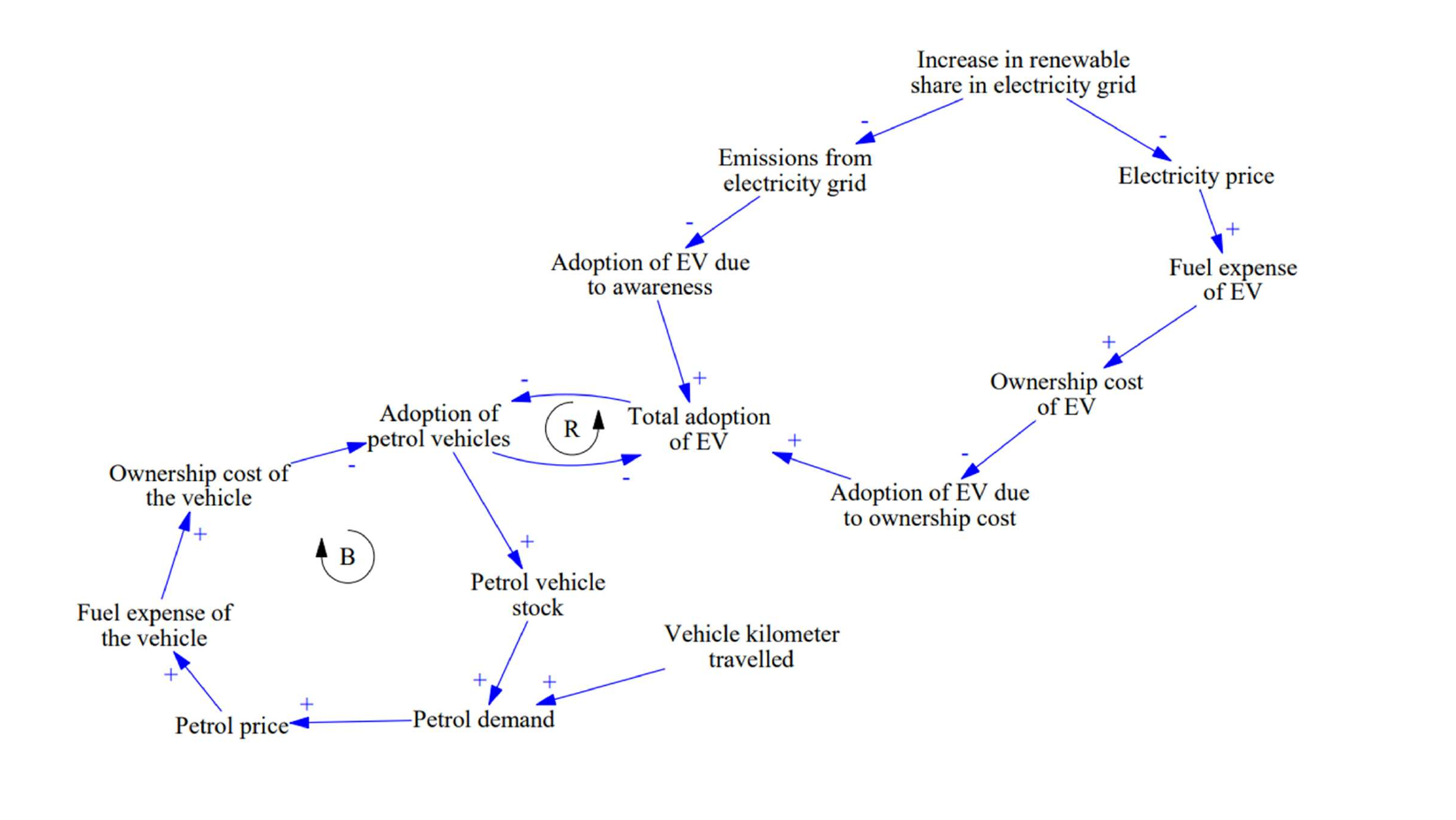}
    \caption{Impact of greener electricity grid and reduced vehicle driving on EV adoption}
    \label{feedback3}
\end{figure}
\begin{figure}[!ht]
    \centering
    \includegraphics[width=17cm,height=17cm, keepaspectratio]{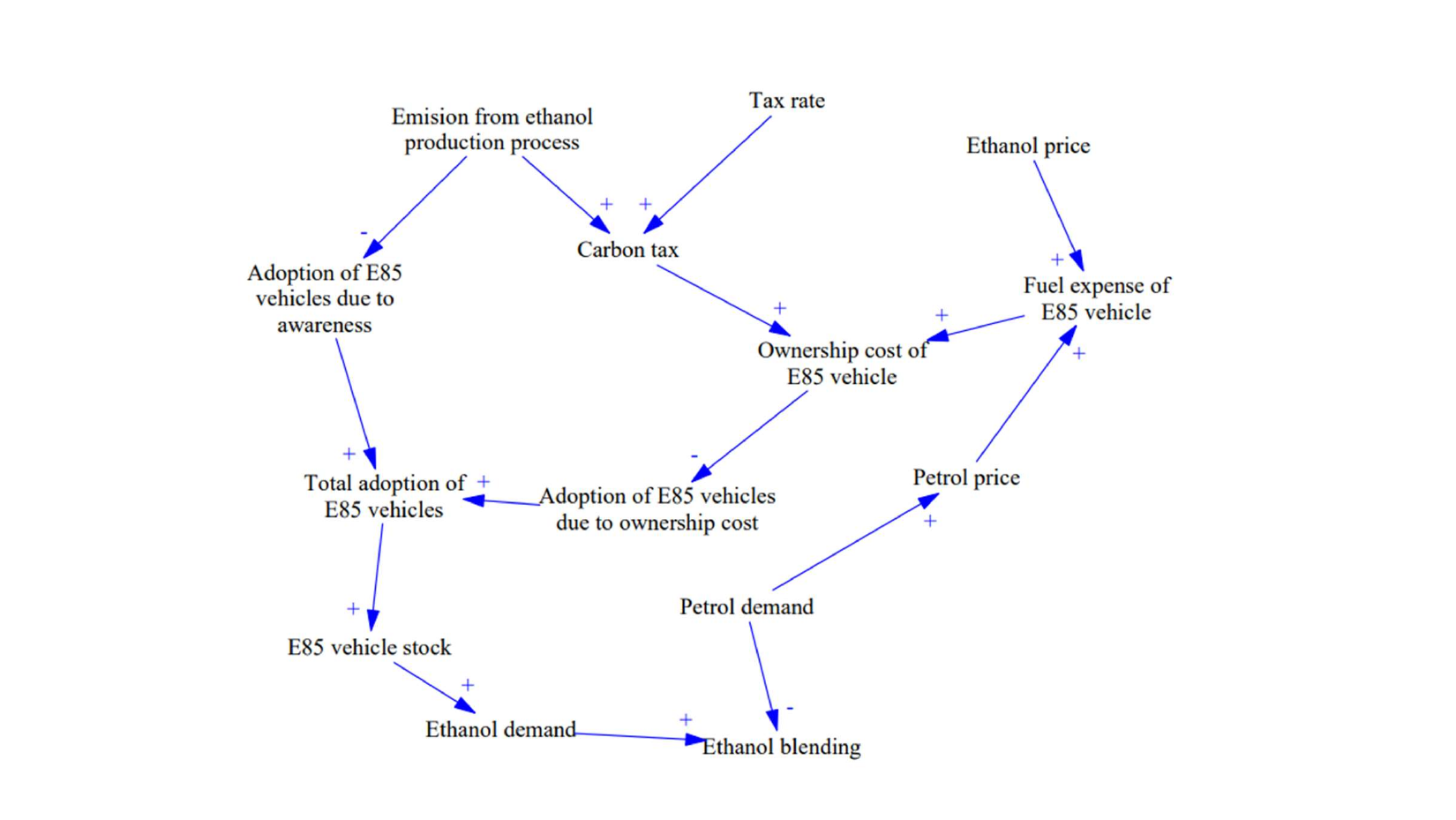}
    \caption{Impact of carbon tax on ethanol blending}
    \label{feedback4}
\end{figure}





% \end{thebibliography}
\end{document}